\documentclass[a4paper,11pt]{article}
\pdfoutput=1 

\usepackage{jheppub} 
\usepackage[export]{adjustbox}
\usepackage[utf8]{inputenc}
\usepackage{multirow}
\usepackage{bbold}
\usepackage[table]{xcolor}
\usepackage{slashed}
\usepackage{bm}
\usepackage{subfig}
\usepackage{float}
\usepackage{fancyvrb}
\usepackage{fvextra}
\usepackage{soul}
\usepackage[title]{appendix}

\usepackage{comment}
\usepackage{tikz}
\usepackage{tkz-euclide}
\usetikzlibrary{shapes.misc}
\usetikzlibrary{decorations.pathmorphing}	
\tikzset{
    v/.style={decorate, decoration={snake, segment length=3mm, amplitude=0.75mm}, draw},
    f/.style={draw=black, postaction={decorate},
        decoration={markings,mark=at position .6 with {\arrow[very thick]{latex}}}},
    fb/.style={draw=black, postaction={decorate},
        decoration={markings,mark=at position .4 with {\arrowreversed[very thick]{latex}}}},
    fnar/.style={draw=black},
    g/.style={decorate, draw=black,
        decoration={coil,amplitude=3pt, segment length=3.5pt}},
    s/.style={dashed,draw=black, postaction={decorate},
        decoration={markings,mark=at position .55 with {\arrow[very thick]{latex}}}},
    sb/.style={dashed,draw=black, postaction={decorate},
        decoration={markings,mark=at position .55 with {\arrowreversed[draw=black,very thick]{latex}}}},
    snar/.style={dashed,draw=black,line width =1.25pt},
    cross/.style={cross out, draw=black, minimum size=2*(#1-\pgflinewidth), inner sep=0pt, outer sep=0pt},
cross/.default={3pt},
}
\usepackage{pifont}

\graphicspath{{Figures/}} 

\def\block(#1,#2)#3{\multicolumn{#2}{c}{\multirow{#1}{*}{$ #3 $}}}

\def\be{\begin{equation}}
\def\ee{\end{equation}}
\newcommand{\ba}{\begin{array}}
\newcommand{\ea}{\end{array}}
\def\bV{\begin{Verbatim}[breaklines=true, breakanywhere=true]}
\def\eV{\end{Verbatim}}

\def\tr{\mathrm{Tr}}

\def\suX{SU(2)_{f}}

\def\Af{V}
\def\Wt{V}
\def\Wtp{{V_p}{}}
\def\Wtm{{V_m}{}}
\def\Zt{{V_{3}}{}}
\def\CKM{V_{\rm CKM}}

\def\lamCKM{\lambda_{{\rm CKM}}}

\def\Leff{\mathcal{L}_{\rm eff}}

\def\thLetau{\theta_{\ell13}}
\def\thLemu{\theta_{\ell12}}
\def\thLmutau{\theta_{\ell23}}

\def\thQdb{\theta_{Q13}}
\def\thQds{\theta_{Q12}}
\def\thQsb{\theta_{Q23}}

\newcommand{\amc}{{\sc MadGraph5}\_a{\sc MC@NLO}}
\newcommand{\fr}{{\sc Feyn\-Rules}}
\newcommand{\superiso}{\texttt{SuperIso}}
\newcommand{\bsmart}{{\sc BSMArt}}
\newcommand{\multinest}{{\sc MultiNest}}

\title{Gauge $SU(2)_f$ flavour transfers}

\author[a]{Luc Darm\'e,}
\emailAdd{l.darme@ip2i.in2p3.fr}
\author[a,b]{Aldo Deandrea}
\emailAdd{deandrea@ip2i.in2p3.fr}
\author[a,c]{Farvah Mahmoudi}
\emailAdd{nazila@cern.ch}

\affiliation[a]{Universit\'e Claude Bernard Lyon 1, 
Institut de Physique des 2 Infinis de Lyon, \\ 
CNRS/IN2P3, UMR 5822, F-69622, Villeurbanne, France}
\affiliation[b]{Department of Physics, University of Johannesburg, 
PO Box 524, Auckland Park 2006,\\ South Africa}
\affiliation[c]{Theoretical Physics Department, CERN, CH-1211 Geneva 23, Switzerland}

\abstract{We introduce the idea of flavour transfer from a non-abelian horizontal $SU(2)_f$ flavour gauge group embedded in the Standard Model flavour structure.
The new flavour vector bosons, in the mass range from the tens of GeV to multi-TeV, do not induce large flavour-changing neutral currents and meson oscillations, which usually provide the dominant constraints on this type of structure. Instead, the main constraints arise from ``flavour-transfer'' operators that we will study in detail.
Several explicit models are presented and their prospects are thoroughly explored, including their phenomenology in the lepton and quark sectors at colliders and lower energy experiments. 
We perform a complete numerical fit in one such scenario, showing that LHC-based lepton-flavour violating searches are competitive with intensity frontier observables.}


\begin{document} 
\hfill {\tt  CERN-TH-2023-139} 
\maketitle

\section{Introduction}

A trademark of the Standard Model (SM) is the scarcity of simple extensions of its gauge group which remain anomaly free in the absence of additional fermions. Apart from the peculiar case of the $B-L$ abelian gauge group in the presence of right-handed neutrinos, all other abelian examples must rely on flavour non-universal charges~\cite{Ellis:2017nrp,Allanach:2018vjg}. The case of non-abelian gauge groups is even more restrictive, as only one  representation acting on the flavour generations may be selected by type of fermion. 

The simplest example relies on introducing an extra horizontal $SU(2)_f$ of flavour, whose global equivalent was repeatedly used as the building block for the Yukawa matrices textures (for some recent works, see for instance~\cite{Linster:2018avp,Fedele:2020fvh,Arias-Aragon:2020bzy}). 
Such a gauge group was suggested very early on~\cite{Monich:1980rr,Berezhiani:1982rr}. Further examples of horizontal non-abelian gauge group were discussed in supersymmetric models (see for example \cite{Berezhiani:1996ii}),

non-supersymmetric models~\cite{Berezhiani:1983hm,Berezhiani:1985in,Berezhiani:1990wn,Barbieri:1996ww,King:2003rf,Grinstein:2010ve} 
and in a few more recent works~\cite{Chiang:2017vcl,Guadagnoli:2018ojc,Belfatto:2018cfo,Belfatto:2019swo,Carvunis:2020exc,Darme:2022uzl}.
Anchoring such SM extensions to energy scales accessible in collider experiments is non-trivial due to the possibility that their breaking -- along with the generation of the hierarchical pattern in the Higgs Yukawa coupling -- could essentially fully decouple from electroweak scale (this issue is not limited to non-abelian cases, see for instance~\cite{Binetruy:1994ru,Binetruy:1996xk,Ibanez:1994ig}).
While the $\suX$ gauge group cannot serve as the only mechanism defining the texture of the Yukawa matrices, it is typically an important ingredient of these larger constructions with a distinct phenomenology from the one of abelian flavour gauge group.
Furthermore, the mass of the horizontal gauge bosons, while tied to the breaking scale, depends on the strength of the horizontal gauge coupling which does not have a direct impact on the texture of the mass matrix. Small new gauge couplings, which would naturally lead to gauge bosons significantly lighter than the breaking scale, are allowed and may push these particles within the accessible energy range at colliders or even in intensity frontier experiments. In fact, we will show that for gauge couplings at the percent level and below, LHC and future HL-LHC searches can adequately provide the strongest limits on these scenarios, thus directly probing the possible NP origin of the SM flavour structure. 
\vskip 0.3cm

In practice, we introduce in this work the idea of flavour transfer and take a closer look at the phenomenology of extensions of the SM gauge group in the case of new flavour $\suX$ gauge symmetries with flavour vector boson masses from the tens of GeV to the multi-TeV range. 
Remarkably, the non-abelian nature of the flavour gauge group provides a high degree of protection against Flavour Changing Neutral Currents (FCNC) and meson oscillations, which usually provide the dominant constraints on this type of new physics (NP) structures.  
Indicating flavour-violating transitions by a $\Delta F$ value,  four-fermion operators originating from flavour gauge boson exchanges will typically satisfy a null sum rule $\Delta F_f +\Delta F_{f'}=0$. This indicates that it is more appropriate to think of their dominant effect as a ``flavour transfer'' within fermions of flavour multiplets rather than a generic flavour violation.
The fundamental reason behind the appearance of these operators stems from the fact that the supposed gauge-symmetric nature of flavour imposes anomaly cancellation restrictions on the $\suX \otimes SU(2)_L \otimes U(1)_Y$ mixed anomalies. This implies that flavour violations in different sectors are always pairwise related, leading to a flavour transfer between two fermion sectors, with specific suppression factors or SM screening, as explained in detail in section \ref{sec:flavtrans}. This link among the different fermions implicitly assumes that anomalies are canceled within the Standard Model without introducing new heavy chiral fermions. This is a minimal and elegant assumption which also avoids stringent bounds from present collider data. This point is discussed in more detail in Sec.~\ref{sec:SU2theo}.

We will show that the  mechanism of flavour transfer predicts clear patterns of flavour-changing processes, governed by the $\suX$-breaking spurions as encoded in the fermion rotation matrices, which can be tested both in intensity frontier experiments and at LHC. We identify several promising intensity frontier observables, and show that in many cases, such as $K_L \to e \mu$~\cite{BNL:1998apv} or $\mu \to e$ transition in matter~\cite{SINDRUMII:2006dvw}, decades-old limits still provide very good constraints and future prospects are bright, while for others, as for instance $K^+ \to \pi^+ \nu \nu$, current searches~\cite{NA62:2021zjw} are actively probing the relevant parameter space. On the other hand, usual lepton-flavour universality violating observables, such as the recent $R_K$ or $R_{K^*}$ measurements by the LHCb collaboration~\cite{LHCb:2022qnv,LHCb:2022zom}, are only generated from $\suX$ breaking terms and correspond to second order effects. We use the code \superiso~\cite{Mahmoudi:2007vz,Mahmoudi:2008tp,Mahmoudi:2009zz,Neshatpour:2021nbn} to calculate a large range of intensity frontier observables. Finally, we explore the potential of LHC and future proton collider experiments in searching for this type of new physics. In particular, the combination of large production rates due to the flavourful content of the proton's quark sea at low energy and of low background thanks to the Lepton-Flavour Violating (LFV) final states, enables collider constraints to compete with -- and in several cases overcome -- intensity frontier limits.

The remainder of this paper is structured as follows. In Sec.~\ref{sec:SU2theo} we introduce the theoretical construction of a $SU(2)_f$ gauge symmetry, and discuss in detail the flavour transfer idea and the properties of the interactions in the mass basis. In Sec.~\ref{sec:LHC}, we discuss and combine the present collider observables in the cases where the SM fermions are chosen to be approximately aligned along specific flavour directions. In Sec.~\ref{sec:flavourobs} we study more specifically the most constrained case of the left-handed model, and analyse in detail the limits coming from flavour observables. We conclude in Sec.~\ref{sec:conclusion} and leave to the appendices more specific material for the interested reader.

\section{Gauge $\suX$ flavour transfers}
\label{sec:SU2theo}
The presence of three fermion families with the same quantum numbers in the SM except for the flavour labels implies the invariance  under a global $U(3)^5$ symmetry, where each $U(3)$ factor mixes family members with identical gauge quantum numbers. This maximal global flavour symmetry of the SM is explicitly broken by the SM Yukawa interactions to the four SM accidental $U(1)$ symmetries which are the three lepton numbers and the baryon number.

In extensions of the SM, this symmetry structure can be modified by the presence of new fields and interactions. Moreover, flavour symmetries in beyond the SM (BSM) theories can be both global and local. However the more general question of flavour in BSM depends not only on the phenomenology directly linked to flavour physics, but is also strongly related to all other aspects touching the model-building hypotheses such as the fundamental or composite nature of the Higgs boson (see for example \cite{Davighi:2023iks}), higher energy symmetries such as unification or supersymmetry, etc. We will consider in this work flavour non-universality at an effective level within a spontaneously broken non-abelian horizontal $\suX$ flavour group embedded in the SM flavour structure. In particular, we study the possibility that a new $\suX$ is part of the flavour symmetry of the SM, in a purely ``horizontal'' gauge fashion. Since there are only three generations, we can embed each type of SM fermions either partially as a doublet, or fully as a $\suX$ triplet. The gauge interactions of $\suX$ are thus totally fixed by the choice of representation type for each SM multiplet
\begin{align}
    Q^\alpha_L,\; D^\alpha_R,\; U^\alpha_R,\; L^\alpha_L,\; E^\alpha_R, \ 
\end{align}
with $\alpha$ the generation (flavour) index.
The flavour structure of the theory depends on the Yukawa part of the Lagrangian, which must lead to
the observed fermion masses as well as the Cabibbo-Kobayashi-Maskawa (CKM) matrix structure. Assuming that the scalars responsible for the breaking of the horizontal gauge groups decouple from the spectrum along with other fields required by the complete UV theory (for instance vector-like fermions, new leptoquarks fields in case of radiative generations of masses, etc...), we can study the effect on flavour observables for the gauge bosons only, using a simplified approach. Flavour symmetry breaking, when the flavour group is gauged, has, among other constraints, generic bounds from flavour-changing neutral currents, setting limits on the mass and couplings of the scalar sector of the theory. 
In the following, we will thus focus on the gauge sector whose phenomenology can be studied relatively separately from the scalar sector and is mostly fixed by the gauge structure itself.
A key requirement to safely decouple the details of the breaking of the gauge groups and the generation of the SM Yukawa couplings from the physics mediated by the gauge bosons is that the low-energy simplified theory is anomaly-free. Otherwise -- among various effects --longitudinal modes corresponding to the Goldstone bosons would be strongly enhanced (see e.g. the recent~\cite{Dror:2017nsg,Dror:2017ehi,DiLuzio:2022ziu,Filoche:2022dxl}, or simply the $m_t^2/m_W^2$ enhancements of loop-induced flavour-violating processes in the SM).

Since new $SU(2)$ gauge groups only introduce $\suX \otimes \suX \otimes U(1)_Y$ mixed anomalies, this leads to the requirement:
\begin{align}
\label{eq:mixedanomaly}
    \left[ C(Q_i) -  C(L_i) \right] - \left[ 2 C(u_{R,i}) -  C(d_{R,i}) - C(e_{R i})\right]  = 0 \ ,
\end{align}
where $C(f)$ is the Casimir of the corresponding fermion $\suX$ representation and we have made explicit the simple cancellations between quarks and leptons which are already at play in the SM. It is clear that this is a rather weak constraint compared to the new Abelian gauge group case, due to the absence of pure $U(1)^3$ and mixed gravity anomalies by construction. 
The Casimir for the doublet and triplet representations of $\suX$ are respectively given by $3/4$ and $2$, for which one can check from Eq.~\eqref{eq:mixedanomaly} that the contributions of doublet and triplet representations must cancel among themselves.\footnote{We do not include right-handed neutrinos in the theory, although it might be attractive to do so in full theories, in particular to escape the presence of the Witten anomaly for scenarios with an odd number of new $\suX$ doublets~\cite{Witten:1982fp}.} In general, we can thus classify the solutions as
\begin{itemize}
    \item (LH) Left-handed, with $C(F_R)=0$
    \item (RH) Right-handed, with $C(F_L)=0$
    \item (B) Baryonic, with $C(L) = C(e_R) = 0$, $C (Q) = C (d_R) = C (u_R) \neq 0$
    \item (L) Leptonic, with $C(Q) = C (d_R) = C (u_R) = 0$ , $C(L) = C(e_R) \neq 0$
      \item (M1) Mixed 1, with $C (Q) = C (e_R) = C (u_R) = 0$ , $C(L) = C(d_R) \neq 0$
    \item (M2) Mixed 2, with $C(L) = C(d_R) =0$ , $C (Q) = C (e_R) = C (u_R) \neq 0$ \, ,
\end{itemize}
where $F_R$ and $F_L$ indicate respectively all right- and left-handed fermions (both quarks and leptons).
These scenarios can be combined two-by-two and mixed between triplet and doublet representations, resulting in $20$ possible new $\suX$ symmetries which can be added to the Standard Model without introducing new chiral fermions for anomaly cancellations.  The main feature of these anomaly-free models will be that several fermionic sector shares the same gauge group, as this will trigger the emergence of the flavour-transfer operators described in the rest of this work. Finally, they can be embedded directly within popular GUT framework, with for instance the (LH) and (RH) scenario corresponding to a Pati-Salam unification gauge group, while the (M1) and (M2) scenario correspond to the usual matter distribution in $SO(10)$ multiplets.

We will focus on the case of doublet representations in the following, leaving the treatment of triplets to future works. Additionally, both fundamental and anti-fundamental representations are isomorphic for $SU(2)$ so that we will concentrate on the former.

\subsection{Gauge sector and residual global symmetries}
\label{sec:gaugesector}
As discussed above, the SM has a full $U(3)^5$ global symmetry in the absence of Yukawa interactions. The presence of new $\suX$ flavour gauge symmetries implies that the new gauge sector will break this large flavour symmetry to a smaller subset depending on the fermions participating in these new interactions. For instance, in the case of scenario (LH), the non-abelian component of the global symmetries in the $Q_L$ and $L_L$ sectors are reduced to two global $SU(2)$ in the presence of the unbroken flavour gauge group, with a linear combination promoted to a local symmetry. The presence -- or absence -- of these global symmetries in the final flavour broken phase then depends on the details of the symmetry-breaking sector. In this section, we shall explore these aspects in detail.
 
\paragraph{Gauge boson masses}

The mass structure of the gauge bosons can be obtained once the scalar sector of the theory is specified. We shall explore the case of scalar fields in both a doublet and/or an adjoint representation of $SU(2)_f$. Writing $\Phi$ for the doublet and $\Sigma$ for the adjoint, the gauge mass terms can be obtained as:
\begin{align}
\label{eq:massWtilde}
\mathcal{L} \supset g_f^2 ~\tr~ &\left[ \Phi^+ (\tau^\dagger \cdot \Wt) (\tau \cdot \Wt) \Phi \right] \ , 
\end{align}
where we have used the $SU(2)_f$ generators $\tau_a ~\equiv~ \sigma_a / 2$, written the $SU(2)_f$ gauge bosons as $\Wt_a$, denoted by $\cdot$ the contraction over the ``$a$'' group indices and written $g_f$ the new gauge coupling. Note that all objects within the trace are thus $2\times 2$ matrices except for $\Phi$. Moving to the Cartan basis, we can write as usual:
\begin{align}
    \tau \cdot \Wt = \tau_+ \Wtp +  \tau_- \Wtm +  \tau_3 \Zt \ ,
\end{align}
with the standard definition for the raising and lowering operators and the corresponding gauge bosons: 
$$\tau_\pm = (\tau_1 \pm i \tau_2) \ ,$$ $$\Wt_{p,m} = (\Wt_1 \mp i \, \Wt_2)/\sqrt{2}\ .$$ Note that we did not use the $\pm$ notation for the $\Wt_{p,m}$ gauge bosons to underline the fact that they are fully neutral under all of the SM charges. The algebra of these operators leads to:
\begin{align}
        (\tau \cdot \Wt)  (\tau^\dagger \cdot \Wt) = \frac{1}{4}( 2 \Wt_+ \Wt_- + \Wt_3 \Wt_3) \times \textrm{Id} \ ,
\end{align}
which together with Eq.~\eqref{eq:massWtilde} make it clear that the mass terms enjoy a global $SU(2)_G$ symmetry when generated by any sets of scalars in the doublet representations of $\suX$.

In practice, for the case of any $\phi^i$ in the fundamental representation (regardless of the VEV orientation), all gauge bosons get the same mass:
\begin{align}
    M_{\Wt_1}^2 = M_{\Wt_2}^2 = M_{\Wt_3}^2 = \frac{g_f^2}{4} \sum_i v_\phi^2 \ .
\end{align}
In the above case, one can expect that the mass degeneracy will be broken by fermionic loops, although potentially only proportionally to small ``spurions'' which will parameterise the symmetry breaking from the mass sector of the theory.
If one is interested instead in a more complex scenario with a significant mixing between the different vector bosons, then a rotation matrix $U_{a}^b$ must be introduced to diagonalise the $\suX$ gauge sector. 

As a side-comment, in the case of a scalar $\Sigma$ in the adjoint representation $\Sigma$. A residual $U(1)$ is preserved so that only two gauge bosons acquire a mass. Using the $SU(2)$ invariance to rotate the VEV to the third component only:
\begin{align}
 \Sigma  \to  \sigma^3_{ij} v_\Sigma \ ,
\end{align}
we get 
\begin{align}
   M_{\Wtp}^2 = M_{\Wtm}^2 = \left(\frac{g_D v_\Sigma}{\sqrt{2}}\right)^2 \ \ ,  \quad M_{\Wt_3}^2 = 0 \ .
\end{align}
Note that this does not imply that the final mass eigenstates will couple in a flavour-preserving way (even if $\sigma_3$ is diagonal), since the fermion mass matrix will depend on the Yukawa sector in a full theory. Nevertheless, we see from these two cases that it is possible to obtain a mass splitting between the gauge bosons, but that the $\Zt$ component will always be lighter than the $\Wt_{p,m}$ bosons.\footnote{This is in line with the results from~\cite{Monich:1980rr}, but seems to run opposite to the simplified model scenario considered phenomenologically in~\cite{Guadagnoli:2018ojc}.} The global residual symmetry corresponds in practice to a ``flavour charge'' for $\Wtm$ and $\Wtp$ and enforces that their masses remain equal. 

In the following, we will restrict ourselves to the case of fundamental representations and $M_{\Wtp}^2 = M_{\Wtm}^2 = M_{\Zt}^2$, and we use $\Wtm$, $\Wtp$ and $\Zt$ as the final mass eigenstates in the gauge sector.

\paragraph{Gauge interactions}

The complete interactions for the flavour bosons $\Wt^a$ with a pair of gauge eigenstate fermions $f_i, f_j$ transforming under the doublet representation of $\suX$ can be obtained by expanding the covariant derivative as:
\begin{align}
\label{eq:gaugefermion}
\mathcal{L}_{int} = - g_f \Wt_{\mu , a } \, \bar{f}^i \, (T^a_{i}{}^{j} \gamma^\mu P_L) \, f_j \ , 
\end{align}
 where the fermions are written in their gauge eigenstates\footnote{Note that this choice is arbitrary, since the SM gauge sector does not distinguish between generations. Any redefinition of the fields will modify only the Yukawa interactions' part of the Lagrangian, which will be only specified in a complete theory of flavour.} and $T^a$ are the generators of $\suX$ in the space of the three fermion generations which can be written by blocks as a function of the Pauli matrices as:
\begin{align}
 T_a ~\equiv~ 
 \frac{1}{2}
 \begin{pmatrix}
\block(2,2){\mbox{$\sigma_a$}} & 0 \\
&& 0 \\
  0& 0 & 0 
\end{pmatrix} \ .
\end{align}
 Note that if one had a mixing between the different vector bosons corresponding to a rotation matrix $U_{a}^b$, the generators for the mass eigenstates must be obtained by  $T_a \rightarrow U_{a}^b  T_b$.
 
Once the gauge interactions to the SM fermions are added, the gauge and fermionic sectors altogether enjoy the $SU(2)$ global symmetry identified previously, which acts on the fermions $F_i$ of a $SU(2)_f$ doublet as either:
\begin{align}
    F_i \to \exp \left( \alpha^a \frac{\sigma_a}{2}  \right)_i^j F_j \quad \textrm{  for $SU(2)_G$ } 
    \end{align}
with the gauge bosons transforming under the standard adjoint representation. The invariance of the interaction term in Eq.~\eqref{eq:gaugefermion} is then automatically satisfied. This symmetry will play an important role in protecting the theory against flavour-changing processes.
 
In order to obtain the gauge interactions in the mass basis, we finally parameterise the diagonalisation matrices of the Standard Model fermions as: 
\begin{align}
\label{eq:su2decomp}
    V_f = \Phi^y V_x V_y V_z,
\end{align}
with the matrix $V_x$ written as:
\begin{align}
\label{eq:paramV}
&V_x=\begin{pmatrix}
1&0&0 \cr 
0 &  e^{i\gamma_x} \cos\alpha_x & -\sin\alpha_x  e^{-i\beta_x} \cr
 0 & \sin\alpha_x e^{i\beta_x} & \cos\alpha_x  e^{-i\gamma_x}
 \end{pmatrix},     
 \end{align}
and $V_y$, $V_z$ defined by rotating instead the $1-3$ or $1-2$ generations. The phase matrix is given by
 \begin{align}
\Phi^y=\begin{pmatrix}
e^{i\phi_{f2}}  &0&0 \cr 
0 &  e^{- i (\phi_{f2} + \phi_{f1})}& 0\cr
 0 & 0 & e^{i\phi_{f2}}   
 \end{pmatrix}.
 \end{align}
 Note that the angles $\alpha_z $ and $\alpha_x $ can be varied in $[-\pi , \pi]$ while the last angle only spans $[-\pi/2, \pi/2]$ (see e.g. the summary in~\cite{Darme:2022uzl}).
 
In general, we can at least absorb one global phase for the quarks and one for the leptons. In the presence of the global  $SU(2)_G$ symmetry, we can then absorb a complete $SU(2)$ factor as well, so that we can parameterise in general the new gauge interaction for at least one fermion sector using the simpler matrix:
\begin{align}
\label{eq:RotationStruct}
    V_{dL} &= \Phi^y (\phi_{f2} = 0) V_{x}  V_{y}  \ .
\end{align}

When the flavour gauge acts on down quarks (thus for all scenarios but the (L) case) we choose to parametrise the mass eigenstates in the quark sector based on the down quarks (so-called ``down basis''). 
The other rotation matrices should be in general parametrised as in Eq.~\eqref{eq:su2decomp}. In particular, for the charged lepton sector, the rotation angle associated with $V_z$ will be of order $1$ in general since it is not protected by any symmetry in our construction. In practice however, the left-handed up quark matrix can be fixed by the measured CKM matrix, and for the left-handed leptons, the neutrino rotation matrix is traditionally chosen equal to the charge leptons' one. Finally, as we are going to see in the next section, each set of effective four-fermion operators in a given fermion sector (e.g. with only charged leptons $\ell_L$) will benefit from a global $SU(2)$ symmetry, so that the additional angles beyond the parametrisations of the form of Eq.~\eqref{eq:RotationStruct} will be only relevant for mixed operators. 
For example in the (LH) scenario that we will explore in full detail in Sec.~\ref{sec:flavourobs}, the angle corresponding to the $V_{z}$ transformation will only appear in mixed $dd\ell\ell$ and $uu \ell\ell$ operators.

For the practical results and approximations of this work, we will focus on the case of real matrices and set the phases to zero, with the matrices $V$ then becoming rotations. Unless stated otherwise, the generic formula will however retain the possible presence of complex phases. 

As an example, in the case of the  (LH) scenario,  the full interaction Lagrangian rotated to the mass basis reads:
 \begin{align}
\label{eq:gaugefermionMass}
\mathcal{L}  \supset - g_f \Af_{\mu , a } \, \left( \bar{u}_L Q_{u}^a \gamma^\mu u_L  +\bar{d}_L Q_{d}^a \gamma^\mu d_L  +\bar{\ell}_L Q_{\ell}^a \gamma^\mu \ell_L  + \bar{\nu}_L Q_{\ell}^a \gamma^\mu \nu_L  \right)\ , 
\end{align}
 where the effective charges are defined as:
 \begin{align}
 \label{eq:effcharge}
     Q_{d}^a  &= V_{dL} T^a V_{dL}^\dagger \nonumber \\
     Q_{u}^a  &= \CKM  V_{dL} T^a V_{dL}^\dagger \CKM^\dagger \\
     Q_{\ell}^a  &= V_\ell  T^a V_{\ell}^\dagger \ , \nonumber
 \end{align}
 and all objects are matrices in the $3 \times 3$ flavour space and $V_\ell$ is parametrised by $3$ angles.

As we will see in Sec.~\ref{sec:texture}, the setups where the rotation matrices in the different sectors are approximately diagonal are of particular interest due to the expected reduction of flavour constraints. We thus will refer to these scenarios based on their flavour ``alignment'' in each fermionic sector, with the notation $(\alpha , \beta)_f$ indicating that both generations $f_\alpha$ and $f_\beta$ are part of a $\suX$ doublet before the rotation to the mass basis.
Since the ordering of the matrices in Eq.~\eqref{eq:su2decomp} is relevant, care must be taken when considering the limit of small spurions, $V_{dL} \sim \mathbb{1}$. The order of Eq.~\eqref{eq:su2decomp} corresponds to a flavour alignment $(12)_Q (12)_\ell$ for $V_{dL}, V_{\ell} \sim \mathbb{1}$. For a $(13)$ flavour alignment, the ordering becomes $V_f = \Phi^z V_x V_z V_y$ and for a $(23)$ flavour alignment $V_f = \Phi^z V_y V_z V_x$. Note that this is not relevant for our full numerical scans since Eq.~\eqref{eq:su2decomp}  is, in any case, a complete parameterisation.

 \subsection{Effective operators decomposition}
 \label{sec:effinteractions}

At energy scales significantly below the electroweak scale (and the flavour gauge boson masses) the new interactions of the theory can be described purely in terms of effective operators in the so-called Weak Effective Theory (WET). 
Since we consider flavour observables at typical scales of a few GeV or below, this effective Lagrangian will be the most important ingredient for intensity frontier experiments, and we can write it as:
 \begin{align}
 \label{eq:effVect}
          \Leff \supset -  \sum_{a,f,f^\prime,\alpha \beta \gamma \delta} \frac{g_f^2}{2} \left( \overline{f}^\alpha \, \frac{ Q^a_{f,\alpha}{}^{\beta}}{M_{\Wt^a}} \gamma^\mu   \, f_\beta \right) \left( \overline{f}^{\prime,\gamma } \,  \frac{ Q^a_{f,\gamma }{}^{\delta}}{M_{\Wt^a}} \gamma_\mu   \, f^\prime_\delta \right) \ ,
 \end{align}
where we have made the flavour indices explicit to emphasise that $f$ and $f^\prime$ are vectors in generation space, while $V_f$ and $T^a$ are $3\times3$ matrices. 
We have included a  factor $1/2$ in the above expression since we do not order the different fermions. Note that the effective couplings  automatically satisfy $(Q_{f}^a)_{i}^{j} = (Q_{f}^{\dagger, a})_{i}^{j}$ due to their gauge origin. Thus the effective Lagrangian is also hermitian. In the following and unless specified otherwise, we will adopt the convention:\footnote{Note that several flavour observables will rely on effective operators defined at the level of the effective Hamiltonian, implying an extra $-1$ factor.}
\begin{align}
      \Leff  ~=~ -  \mathcal{H}_{eff}  ~=~  \sum_{O_m = O_m^\dagger} C_m O_m +  \sum_{O_m \neq O_m^\dagger} \left( C_m O_m +  h.c.\right) \ ,
\end{align}
and systematically order the effective operators such that: (1) in mixed operator (down) quark fermion pairs are used first, (2) the heaviest flavour number must be the last one in a fermion pair. Finally, for operators which involve the same type of chiral fermions (e.g. purely leptonic ones), we will use Fierz identities to obtain a non-redundant basis with only operators ordered as $ijkl$ with $ k \geqslant i $ and $l \geqslant j$.

Since $M_{\Wt} = M_{\Wt_1} = M_{\Wt_2} = M_{\Wt_3}$ we can use the completeness relations on the generators of $SU(2)_f$, the effective couplings before rotating to the final mass eigenstates are given by:
 \begin{align}
 \label{eq:effVect1}
     \Leff \supset -  \sum_{a,f,f^\prime} \frac{g_f^2}{8  M^2_\Wt}( 2 \delta^{il} \delta^{jk} - \delta^{ij} \delta^{kl} )\left( \overline{f}_i \gamma^\mu \, f_j \right) \left( \overline{f}^\prime_k \gamma_\mu \, f^\prime_l \right)  \ ,
 \end{align}
where $i,j,k,l$ are $SU(2)_f$ indices. In the mass basis $F_\alpha$, with $\alpha$ an $SU(3)$ flavour index, the effective operators become
 \begin{align}
 \label{eq:effVectMass}
     \Leff  \supset -  \sum_{a,F,F^\prime} \frac{g_f^2}{8 M^2_\Wt}  (2 U_{F F^\prime}^{\alpha  \delta} U_{F F^\prime}^{\dagger, \zeta \beta} -U_{F F^\prime}^{\alpha \beta} U_{F F^\prime}^{\dagger, \zeta \delta})  \ \left( \overline{F}_\alpha \gamma^\mu \, F_\beta \right) \left( \overline{F}^\prime_\zeta \gamma_\mu \, F^\prime_\delta \right)\ , 
 \end{align}
 where we have defined the effective mixing combinations
 \begin{align}
  U_{F F^\prime}^{\alpha \beta} ~
  \equiv~  (V_{F})^{\alpha, \zeta} (V_{F^\prime}^{ \dagger})_{\zeta}^{\, \beta} \ ,
 \end{align}
with $a$ representing the flavour indices of the fermions which are part of the $SU(2)_f$ doublet (thus leaving out one generation). The above effective interactions are thus highly structured, which implies automatically that several combinations will be strongly suppressed. 

Yet, the scalar fields used to break the $\suX$ symmetry are not affected by the global symmetries described above and this implies that the Yukawa operators will induce an explicit breaking which we can use to parameterise the symmetry-breaking effective operators. 
 We fix the down quark diagonalisation matrix as in Eq.~\eqref{eq:RotationStruct}. The up-sector rotation matrix $V_{uL}$ is then fixed phenomenologically by the measured CKM matrix. 
 Altogether, using the decomposition~\eqref{eq:su2decomp} and the global symmetries we obtain the following parameterisation:
\begin{align}
    V_{dL} &= V_{dL x}  V_{dL y}  \ ,\nonumber \\
        V_{uL} &= \CKM V_{dL}  \ ,\\
    V_{\ell} =& V_{\ell x}  V_{\ell y} V_{\ell z}  = V_{\nu}  \ . \nonumber
\end{align}

As a final comment, we point out that if the three gauge boson masses are not equal, but that we instead have $M_{\Wt_{pm}} \neq M_{\Zt}$, the global symmetry preserved in the gauge sector is an abelian  $U(1)_G$ . One can attribute to each part of flavour doublet a ``flavour charge''. Focusing on the case with only one fermion type $f$ in the currents, the structure of the effective operators is given by
 \begin{align}
 \label{eq:effVect2}
     \mathcal{L}_{eff} \supset -  \sum_{f} \left[   \frac{g_f^2}{8 M^2_{\Zt}}  \left( \overline{f}_1 \gamma^\mu \, f_1  - \overline{f}_2 \gamma^\mu \, f_2  \right)^2  + \left( \frac{g_f^2}{4 M^2_{\Wtp}}  \left( \overline{f}_2 \gamma^\mu \, f_1 \right) \left( \overline{f}_1 \gamma_\mu \, f_2 \right)+ h.c.  \right)\right] \ ,
 \end{align}
where the indices again refer to the generation and  $S$ is a symmetry factor as described above.
Nonetheless at the gauge interaction level, a global $U(1)_F$ prevents the appearance of any meson-mixing contribution as it corresponds to charge $2$ operators. This protection was noticed already in early works on the topic of horizontal gauge symmetries, see e.g.~\cite{Monich:1980rr,Berezhiani:1989fs,Berezhiani:1998wt}.
 
The quark mass matrix explicitly breaks the global $U(1)_G$. Using the decomposition Eq.~\eqref{eq:su2decomp}  we obtain:
\begin{align}
    V_{dL} &= V_{dL x}  V_{dL y}  V_{dL z} \,\nonumber \\
    V_{uL} &= \CKM V_{dL} \ .
\end{align}
In order to prevent too large oscillations for the $K$ and $D^0$ mesons, these rotations must be as close to identity as possible,  in particular for the one acting on the 1-2 sector $ V_{dL z}$. As was noticed in~\cite{Guadagnoli:2018ojc}, even enforcing directly  $ V_{dL z} = \mathbb{1}$ is not enough since the CKM rotation leads in the up-quark sector to an off-diagonal term proportional to $\lambda^2$. Additionally, this  case also corresponds more closely to the usual model of individual $Z^\prime$ new boson with flavourful interactions, for which there exists already ample literature. 

In the following we stress that we concentrate on the $SU(2)_G$ case with $M_{\Wt} = M_{\Wt_1} = M_{\Wt_2} = M_{\Wt_3}$.

\subsection{Textures and the mass-mixing hierachy problem}
\label{sec:texture}

In proceeding to study the phenomenology of the new gauge bosons, the mass rotation matrices will be key inputs. 
 In general, one expects the new gauge symmetry to be broken at a relatively high scale as the flavour-violating processes must be sufficiently suppressed.  
Furthermore, it is clear that while the $\suX$ gauge group described in the previous section is an important ingredient in the structure in these matrices, it cannot be the only one. 
Indeed, taking as an example the (M1) scenario and focusing on the down quark sector with $(12)$ flavour alignement, the mass matrix can be expressed in the schematic triangular form (after a $U(3)_{Q_L}$ re-parameterisation of the right-handed fermions)
\begin{align}
    \mathcal{M}_d \sim \begin{pmatrix}
    \epsilon & 0 & 0 \\
    \epsilon & \epsilon & 0 \\
     \epsilon &  \epsilon & 1 
    \end{pmatrix} \ .
\end{align}
In particular, we do not distinguish between two of the fermionic generations (in the example above, the first and second). This generic feature stems from the fact that all $\suX$ spurions $Y_i$ are a priori accompanied by the dual spurion $\epsilon_{ij} Y^{\dagger ,j}$, effectively preventing an additional hierarchy.
Thus we expect an additional structure in the UV to generate the full fermion structure. In the following we present a short review of existing UV-complete scenarios that will be impacted by the constraints studied via our low-energy $\suX$ case.

A first set of top-down setups which naturally encompass the $SU(2)$ scenario discussed in this work as a low-energy theory are based on the larger flavour gauge group $SU(3)$. This group is typically broken at a high scale to a $SU(3) \to SU(2) \times U(1)$ pattern, with the $U(1)$ gauge group having non-universal gauge couplings and the $SU(2)$ a non-abelian sub-group similar to those studied in this work. Earlier works pre-dating the top-mass measurements typically focused on Grand-Unified theories, such as a left-right model~\cite{Berezhiani:1983hm}, $SU(5)$~\cite{Berezhiani:1982rr,Berezhiani:1985in,Berezhiani:1990wn} or Pati-Salam constructions~\cite{King:2003rf} (this last work also introduces various additional discrete symmetries, including a $Z_5 \times Z_2 \times Z_2$ one). The corresponding $SU(2)$ sub-groups are often naturally matching the anomaly-free models described above: $SU(5)$ unification group corresponding to a (M1) and (M2) scenarios, Pati-Salam models to the (LH) and (RH) scenarios. In all cases, obtaining both the CKM matrices and the fermion hierarchies leads to highly-hierarchical rotation matrices for the fermions. Depending on the approaches taken to generate the $SU(3)$-breaking Yukawa couplings, the resulting $SU(2)$ sub-group will have either a $(23)$ flavour alignment~\cite{Berezhiani:1983hm} or a $(12)$ flavour alignment~\cite{Berezhiani:1982rr,Berezhiani:1985in,Berezhiani:1990wn,King:2003rf}.
Note that the more recent reference~\cite{Grinstein:2010ve} (and subsequent literature focusing on dark matter aspects) follow a similar approach but without the unification gauge groups, thus leading to several horizontal $SU(3)$ gauge groups for the different types of fermions.
In all of the examples above, pseudo-Nambu-Goldstone bosons may be present, this is however strongly dependent on the details of the symmetry-breaking sector (and has been abundantly studied in the literature since early on~\cite{Wilczek:1982rv}, see for instance the recent review~\cite{DiLuzio:2020wdo}).  
Similarly the residual $U(1)$ flavour non-universal gauge group leads to phenomenological consequences that have been abundantly studied, and the mass of its gauge boson depends on the details of the symmetry-breaking sector. We will thus assume in the rest of this work that these particles are either rendered massive or feebly-coupled enough to escape detection and focus on the phenomenological consequences of the non-abelian $\suX$ gauge bosons. In a complete UV model based on the above structures, one should therefore supplement the constraints discussed in the rest of this work with those arising from to the extra particles or interactions.

Finally, an approach put forward more recently in~\cite{Greljo:2023bix,Antusch:2023shi},\footnote{Both references appeared few months after the first version of this work.} does not rely on a larger $SU(3)$ gauge group but introduces only a gauged $SU(2)$ to explain the flavour structure. These models rely instead on the radiative generation of mass of the first generation fermions. Interestingly, the texture structure in~\cite{Greljo:2023bix} corresponds to the (L) scenario described above and predicts highly hierarchical mixing matrices, with off-diagonal elements typically of the order of the ratios of the corresponding fermion masses.

Altogether, the UV models referenced above which incorporate the considered $\suX$ gauge group in their low-energy limits typically predict almost diagonal rotation matrices for the fermionic sectors coupled to $\suX$. It is thus clear that in our bottom-up approach of Eq.~\eqref{eq:effcharge} a particularly important limit is the case where they are approximately diagonal.
Assuming for instance that $\suX$ is orientated toward the two first generations and focusing on the down quark sector, we obtain:
\begin{align}
\label{eq:rotationmat}
     V_{dL x} = \mathbb{1} + \begin{pmatrix}
     0  & 0  &0 \\
     0  & 0  & - \epsilon_{23} \\
     0  & \epsilon_{23}  &0 \\
     \end{pmatrix} \qquad  V_{dL y} = \mathbb{1} + \begin{pmatrix}
     0  & 0  &- \epsilon_{13} \\
     0  & 0  &  \\
     \epsilon_{13}   & 0 &0 \\
     \end{pmatrix} \ .
\end{align}
The effective mixing parameter in the down sector then takes the form:
\begin{align}
\label{eq:coupdL}
     U_{d_L d_L} = \mathbb{1} + \begin{pmatrix}
     - \epsilon_{13}^2  & - \epsilon_{13} \epsilon_{23}   &  -\epsilon_{13} \\
     - \epsilon_{13} \epsilon_{23}   &   - \epsilon_{23}^2   &  -\epsilon_{23}   \\
     -\epsilon_{13} &  -\epsilon_{23}  &  \epsilon_{23}^2 +  \epsilon_{13}^2  \\
     \end{pmatrix} + \mathcal{O} (\epsilon^3) \ .
\end{align}
The 1-2 generation mixing are suppressed by $\epsilon_{13} \epsilon_{23}$ while the 2-3 mixing only by $ \epsilon_{23} $. It is clear that the pattern $\epsilon_{13} 
\ll \epsilon_{23} \ll 1$ will then represent a particularly protected direction.
In the up sector, the approximate unitarity of the CKM matrix in the Cabibbo sub-region leads also to a strong suppression of the flavour-violating off-diagonal terms. In the Wolfenstein parameterisation, we obtain at first order in $\epsilon$ and up to $\lambda^3$ contributions:
\begin{align}
\label{eq:coupuL}
     U_{u_L u_L} = \mathbb{1} + \begin{pmatrix}
    0  &  A \lambda^2 \epsilon_{13}   &  -\epsilon_{13}  \\
     A \lambda^2 \epsilon_{13}   &    2   \epsilon_{23} A \lambda^2  &  -\epsilon_{23}  + A \lambda^2   \\
     -\epsilon_{13}    \ \  &  -\epsilon_{23} + A \lambda^2   &    -2 \epsilon_{23}  A \lambda^2\\
     \end{pmatrix} + \mathcal{O} (\epsilon^2,\lambda^3)\ .
\end{align}
Flavour violation between the first and second generations is still suppressed by at least a power of $\epsilon$.
In particular, Eqs.~\eqref{eq:coupuL} and~\eqref{eq:coupdL} imply that one of the most constraining flavour observables, namely meson mixing, will be very strongly suppressed (at least at the level of $\epsilon_{13} \epsilon_{23} \lambda^3 \eta$ in terms of Wolfenstein parameters).
Note that one could have equally chosen to parameterise first the up quark sector, then use the CKM matrix to obtain the down quark rotation matrices, in which case, the above formula would simply be inverted.

\subsection{Flavour transfer from $\suX$ gauge symmetries}
\label{sec:flavtrans}

As can be seen from Eq.~\eqref{eq:effVect2}, effective operators obtained by integrating out the $\suX$ gauge boson do not lead to flavour violation in each fermionic sector independently.  This can be understood from the presence of the various global $SU(2)$ symmetries associated with each effective four-fermion operator within a fermionic sector. They strongly suppress flavour-violating processes confined to the up, down, or lepton sectors (see e.g. Eqs.~\eqref{eq:coupdL} and \eqref{eq:coupuL}), corresponding to $uuuu$, $dddd$ or $\ell \ell \ell \ell$ processes.

Following the standard notation of classifying the number of flavour-violating transitions as $\Delta F$, we observe that the possible four-fermion operators then satisfy the sum rule:
\begin{align}
\label{eq:flavtransfer}
    \Delta F_f + \Delta F_{f^\prime} = 0 \ .
\end{align}
This illustrates the fact that in the absence of spurions, the $\suX$ gauge interactions can only lead to flavour ``transfer'' between two fermion sectors (for instance, $s \to d \mu^- e^+$ in the (LH) scenario). Due to anomaly cancellation, Eq.~\eqref{eq:mixedanomaly}, imposes that at least two fermion sectors must be charged under one $\suX$. Such transitions are always possible, even in the absence of $\suX$-breaking spurions. This kind of process will likely be at the core of the phenomenology of $\suX$ models.
In fact, as we will illustrate explicitly in the next section for the case of the (LH) scenario, a clear pattern emerges for $\suX$ scenarios: unprotected operators arise from mixed quark/lepton fermionic bilinears: $(\bar u u ) (\bar \ell \ell)$, $(\bar d d ) (\bar \ell \ell)$.
Additionally, while transitions between the components of $SU(2)_L$ doublets (and in particular $uudd$ vertices) are not as protected, they also correspond to the effective interactions mediated by the $W$ boson and thus BSM limits would suffer from a large SM background. Unsuppressed, ``golden'' modes for searches of $\suX$ models can be therefore classified for the scenarios listed at the beginning of Sec.~\ref{sec:SU2theo}.

\begin{itemize}
    \item (L) scenario, with doublets $L_i, e_{R, i}$. The only mixed operators are of the form $\ell \ell \nu \nu$. The $\suX$ interaction for right-handed leptons implies for instance that 
    $$\mu^-_R \to e^-_R \bar{\nu}_e \nu_\mu$$ decays are allowed, with percent-level constraints on the corrections to the Fermi theory available from muon decay constants~\cite{Belfatto:2019swo,Workman:2022ynf}. The superposition of left-handed and right-handed couplings will also imply that radiative processes will be important at one-loop, with relevant channels being:
    \begin{align}
        \mu \to e \gamma \, , \ \tau \to \ell \gamma \, , \ (g-2)_\ell \ .
    \end{align}
    \item (B) scenario, with doublets $Q_{L,i}, u_{R, i}, d_{R, i}$. Similarly to the leptonic case, the only mixed operators involve the same fermions as effective operators in Fermi's theory $u u d d$. Radiative processes will be sensitive to the combination of right-handed and left-handed interactions, including:
        \begin{align}
       B \to K^{(*)} \gamma , B \to \rho \gamma \ .
    \end{align}
    We leave a detailed study of this case, and more precisely of the limits from other purely hadronic flavour transfer processes to future works.\footnote{In particular $SU(2)_f$ gauge symmetry prevents the appearance of a renormalisable kinetic mixing term between the new gauge bosons and the SM $SU(2) \otimes U(1)_Y$ ones. However, after the breaking of both gauge symmetries (which we assumed is dominantly encompassed by the spurions in our case and diagonal masses for the $SU(2)_f$ gauge boson), the loop-induced contribution should lead to a non-zero $\Zt-Z-\Wt$ mixing. We however expect the corresponding loop-induced constraints to be mostly important for the (L) and (B) scenarios due to the weak bounds from flavour transfer processes.} 
    \item (RH) scenario, with doublets $e_{R,i}, u_{R, i}, d_{R, i}$. The presence of couplings to both down quarks and leptons leads to very strong constraints from rare meson decays. Relevant channels depend on the flavour orientation and include:
    \begin{align}
K \to \pi e \mu \, , \quad K_L \to e \mu \\
B \to K e \mu \, , \quad B_s \to e \mu \\
B \to \pi e \mu \, , \quad B_d \to e \mu \ . 
    \end{align}
    In the special case where the lepton and quark flavour alignments are not similar, with a $(12)_Q$ alignment for quarks and $(13)_\ell$ or $(23)_\ell$ for leptons, the kaonic decays become kinematically forbidden, and hadronic tau decays, $\tau \to \ell K^\star, \tau \to \ell K_S $ become dominant. 
    \item (M2) scenario, with doublets $Q_{L,i}, u_{R, i}, e_{R, i}$. As both up and down left-handed quark sectors have $\suX$ doublet, the CKM matrix will force the appearance of a $\Delta F = 1$ operator (for instance $uu e \mu$), in contrast with the previous cases. This implies that in addition to the processes shown above for the (RH) scenario, muon conversion in nuclei
    \begin{align}
        \mu \ \textrm{Au} \to e   \ \textrm{Au} 
    \end{align}
    can play an important role.
    \item (M1) scenario, with doublets $L_{i}, d_{R, i}$. The relevant channels are similar to the ones of the (RH) case. The additional coupling to the neutrino sector means that several missing energy channels become relevant, for instance:
        \begin{align}
K \to \pi \nu \nu \, , \ B \to \pi \nu \nu \, , \ B \to K \nu \nu \ .
    \end{align}
    \item (LH) scenario, with doublets $Q_{L,i}, L_{i}$. This is the most constrained case as it contains all the features of the above, from rare meson decays to muon conversion in nuclei and neutrino final states. We will make an in-depth study of the relevant intensity-frontier constraints in Sec.~\ref{sec:flavourobs}. 
\end{itemize}

In all of the above, it has been assumed that the vector boson mass is significantly higher than the meson mass scales (although it can still be largely below the electroweak scale). In principle, the $\suX$ gauge coupling may be small enough such that $M_V \sim \textrm{GeV} $. In this case, the effective operators described above become momentum-dependent due to the gauge boson propagator leading to terms of the form 
\begin{align}
\frac{M_i^2-s}{\Gamma _i^2 M_i^2+\left(M_i^2-s\right){}^2}\longrightarrow _{s\gg M_i^2} \ \ \frac{1}{s} \ .
\end{align}
While the cancellations described in the previous section apply, the dominant processes will be the \textit{on-shell decay} of mesons via the emission of a light $\Wt$ boson, following $s \to d \Wt, b \to s \Wt$ or even $\mu \to \Wt e$ at lower masses. While nothing a priori prevents this situation from occurring  for couplings small enough to be unconstrained, it clearly corresponds to a different phenomenology altogether that we leave for future work. In the following, we will therefore focus on the case of new  $\suX$ gauge bosons with masses above $10$ GeV, so that all of the flavour physics observables will be described by effective operators.

\section{Collider prospects}
\label{sec:LHC}

The gauge bosons of the $\suX$ flavour gauge group have the experimentally attractive property that in all scenarios but for the (B) and (L) cases, they couple both to quarks and leptons. They can thus be abundantly produced at LHC via the (sea) quarks present in the protons, and be easily detected using their decay into leptons, therefore escaping the large QCD background. Combined with the fact that their decay can be mainly into lepton-flavour-violating (LFV) final states (thus reducing dramatically the QED background), these bosons are a perfect target for both LHC and future hadronic colliders. Collider constraints on both di-jet and di-lepton searches could play a key role in testing these setups, together with the study of LFV resonances. 

\begin{figure}[t!]
\centering
        {\includegraphics[width=0.75\linewidth]{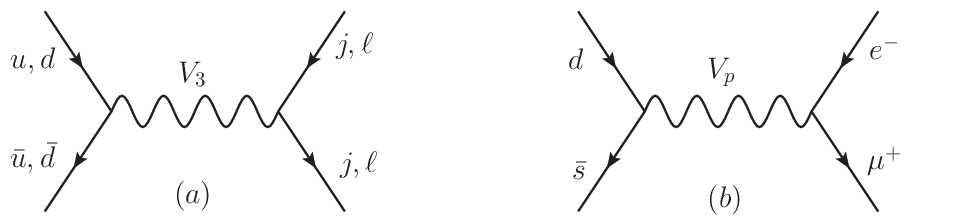}} \hspace{0.2cm}
        \caption{Example of relevant processes for the LHC limits on our $SU(2)_f$ model for both the $\Zt$ and $\Wt$ boson coupling to first and second generations only. }
\label{fig:SU2lquark3}
\end{figure}

In order to explore the collider phenomenology of our $\suX$ models, we have implemented them in the
\fr/UFO~\cite{Christensen:2009jx,Degrande:2011ua,Alloul:2013bka} formalism, then used the platform \amc~\cite{Alwall:2014hca} to generate the cross-section times branching ratio $\sigma \times \mathcal{B}$ for all the processes listed in the next sections (we show two examples of such processes in Fig.~\ref{fig:SU2lquark3}).

\subsection{Same flavour final states} The $\Zt$ boson can be produced by a very large range of processes at the LHC, thanks to its couplings to $u \bar{u}$ and $d \bar d$ for production and its decay into leptonic $e \bar e$, $\mu \bar \mu$ or $\tau \bar \tau$ final states. 

We first include di-lepton resonant searches. Flavour-preserving final states typically arise from an s-channel $\Zt$ boson $pp \to \Zt \to \ell^+ \ell^- $(see Fig.~\ref{fig:SU2lquark3}). At large masses, we rely on the limits on  $\sigma \times \mathcal{B}$ from the CMS analysis~\cite{CMS:2021ctt}, choosing the final states $e^+ e^-$, $\mu^+ \mu^-$ -- or both -- depending on the leptonic flavour alignment.  For the masses around and below the hundred of GeV, we use 
a dedicated CMS search on very narrow resonances~\cite{CMS:2019buh}. We put a limit on $g_f$ up to a few $10^{-3}$ by recasting the reported limit on a dark photon model. We obtain the dark photon $\sigma \times \mathcal{B}$  in \amc~by the same procedure as for the $\suX$ model. These low mass di-muon limits represent a significant achievement and improve former LEP-based searches~\cite{Leike:1998wr} applicable to our scenario by more than an order of magnitude.

In the entire parameter space of the considered models, the new vectors produced from the quark couplings can also decay into a pair of jets. This final state however suffers from a much larger background than the di-lepton one. Thus, di-jet searches will generically lead to weaker constraints than the di-lepton ones for the $\suX$ models with both lepton and quark couplings that we focused on in this work.
Model-dependent limits on a new  $\Zt$ boson with couplings similar to baryon numbers are widely available (see in particular the summary in~\cite{Dobrescu:2021vak}). As a first step, we have thus implemented a B-number gauge boson in \fr/UFO~\cite{Christensen:2009jx,Degrande:2011ua,Alloul:2013bka}, then generated using \amc~~~the cross-section times branching ratio $\sigma \times \mathcal{B}$ for di-jet final states from $pp \to Z^\prime \to j j $. In a second time we have implemented our complete $SU(2)_f$ model, including both $\Zt$ and $\Wtm,\Wtp$ bosons, and used the corresponding  $\sigma \times \mathcal{B}$ to rescale the limits from Refs.~\cite{ATLAS:2019fgd,CMS:2019gwf,ATLAS:2018qto,CMS:2016ltu,ATLAS:2019itm,CMS:2019emo,CMS:2019xai}. At small masses, we need the processes with ISR photon or jets, thus simulating $ p p \to \Wt a$ and $ p p \to \Wt j$ instead. In all cases, we included b-jets in the final states so that this limit applies also for $(13)_Q$ and $(23)_Q$ flavour alignments.

As we will see, this set of limits do not give the dominant constraints on our model but are nonetheless of relevance since they are independent of the leptonic sector and typically cover the coupling range above $g_f \gtrsim 0.02$ for masses between the tens of GeV up to a few TeV.

\subsection{Different flavour final states}

At masses below the TeV, di-lepton searches typically suffer from a strong QED background, which can be significantly alleviated when considering flavour-mixed final states such as $e \mu$, $e\tau$ or $\mu \tau$. 
For invariant masses above a few hundred of GeV, we use the recent CMS analysis~\cite{CMS:2022fsw}.
Since our model predicts narrow resonances in the range accessible by LHC, we use the model-dependent limits derived for an RPV-SUSY sneutrino-like signal down to $200$ GeV, leveraging the cross-section estimate of the \amc~\cite{Alwall:2014hca} for the $\Wt$-induced processes. We use the $e \mu$, $e\tau$ or $\mu \tau$ final states depending on the leptonic flavour alignment.

In all cases, the experimental sensitivity suffers significantly below the TeV due to the large bin sizes compared to the NP particle width and experimental resolution.
At invariant masses around the SM $Z$ and $H$ boson masses, finer-grained analyses have been performed by the ATLAS and CMS collaborations searching for the LFV $Z, H$ decays.
While these searches focused primarily on the SM particles, data can be scouted for the presence of other LFV resonances, with $Z$ searches covering the mass range $[70,110]$ GeV and the $H$ searches covering the range $[110,160]$ GeV ($\tau$ LFV searches typically cover a larger range).
We use both the ATLAS~\cite{ATLAS:2019old} and CMS~\cite{CMS:2023pte} results for $ e \mu$ searches in the Higgs mass window, with in particular the later search being already interpreted in terms of a limit on a NP process (we will discuss at the end of this section the anomaly in this dataset). For $\tau$ LFV processes $e \tau$ and $\mu \tau$ in the Higgs window, we use the CMS analysis~\cite{CMS:2021rsq}.
In the $Z$ window, we use the two dedicated ATLAS analyses~\cite{ATLAS:2022uhq}  ($e \mu$ final state) and~\cite{ATLAS:2021bdj} ($e \tau$ and $\mu \tau$ final state). 

While the recent CMS analysis in the $H$ region does interpret the results in terms of NP constraints, all the other LFV analyses in the $H$ and $Z$ window focus on the SM signal, and their reach must be obtained via a re-interpretation of the public data.
In order to obtain a first estimate of the recasted constraints, we proceeded in steps.

\begin{itemize}
\item We first obtain the signal shape by assuming that the NP signal would mimic the $H$ and $Z$ sample signal reported by the experimental collaborations. In the case of the $Z\to e \mu$, the experimental resolution is small enough so that the $Z$ width plays a role in the reported signal in~\cite{ATLAS:2022uhq}. We thus used a Gaussian calibrated on the $H \to e \mu$ signal~\cite{ATLAS:2019old} to smear the resonance with a width of 
    \begin{align}
    \sigma_{m_{e\mu}} \sim 2.7 \, \textrm{ GeV} \sqrt \frac{m_{e\mu}}{125 \, \textrm{ GeV}} \ ,
    \end{align}
    which led to a satisfactory agreement of the $Z \to e \mu$ signal once added to the $Z$ width. For the $\tau$ final states, we instead directly use the reported signal shape, shifted to the mass of our NP vector states.
    \item  Second, we derive the corresponding $95\%$ C.L. limit from the binned smeared signal found in the first step with the pyhf software~\cite{pyhf,pyhf_joss}. 
    For the $\tau$ LFV final states, the collaborations~\cite{CMS:2021rsq,ATLAS:2021bdj} directly reported the error on the fitted background estimate. The CMS search~\cite{CMS:2021rsq} optimised their analysis strategy via a BDT to the Higgs putative signal and our recast limit based on their public data is significantly weaker, with a dedicated re-interpretation of the analysis pipeline likely leading to stronger constraints. For the $e \mu$ LFV final state in the $Z$ window, the background fit uncertainty is only graphically reported in~\cite{ATLAS:2022uhq}, we thus calibrated it to obtain a limited number of events compatible with the one reported for the $Z \to e \mu$ process. We applied the same approach for the recasting of the $H\to e \mu$ analysis~\cite{ATLAS:2019old}.
    \item Finally, we assume that the NP process would have the same experimental efficiency and acceptance $\epsilon \times \mathcal{A}$ as the SM $H$ and $Z$ ones to obtain the limits on our model using the simulated cross-section from \amc.
\end{itemize}
It is clear from the above discussion that, with the exception of the CMS $e \mu$ LFV search, the limits recast for LFV final states come with a significant theoretical error, and are likely quite conservative (in particular for the $\tau$ LFV processes).

\subsection{Collider and flavour transfer constraints} 

\begin{figure}[t]
\centering
\subfloat[]{%
\includegraphics[width=0.47\textwidth]{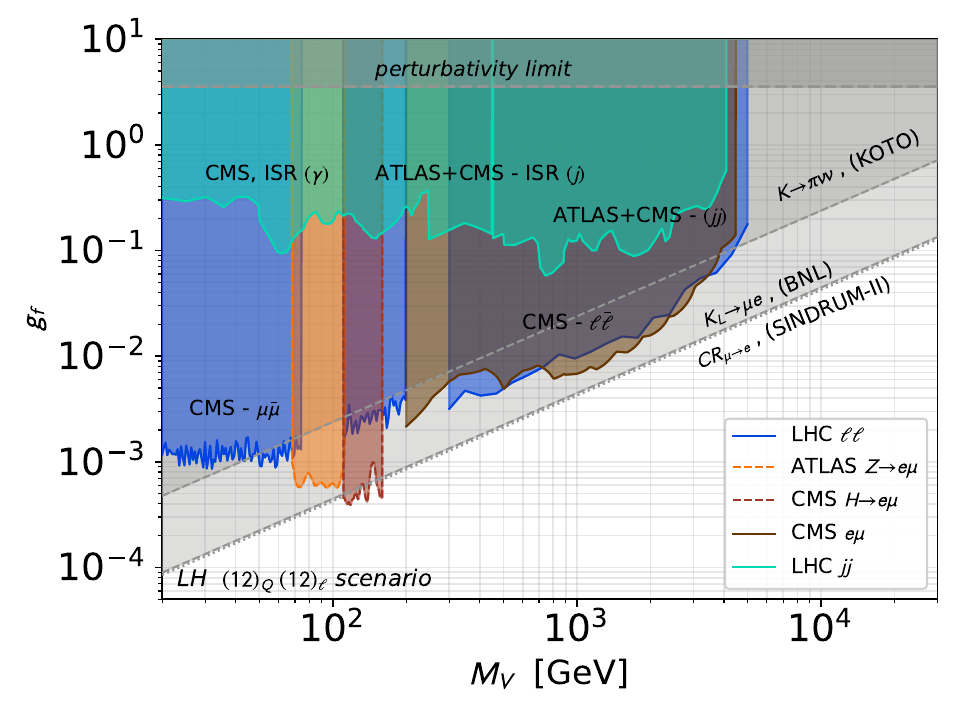}
}%
\hspace{0.02\textwidth}
\subfloat[]{%
\includegraphics[width=0.47\textwidth]{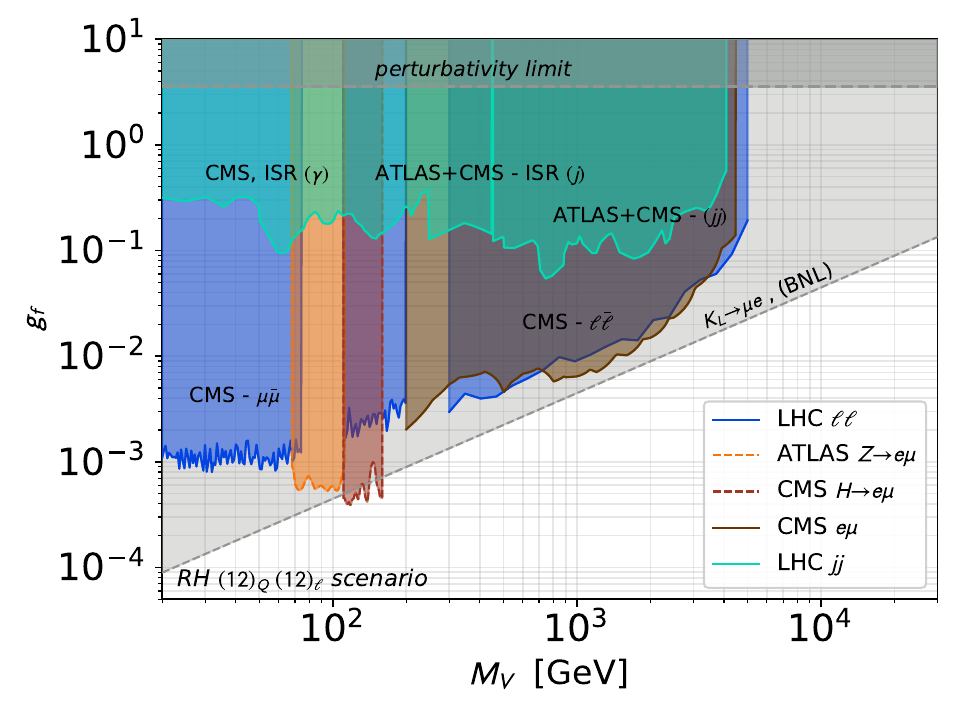}
}%
\hspace{0.02\textwidth}
\subfloat[]{%
\includegraphics[width=0.47\textwidth]{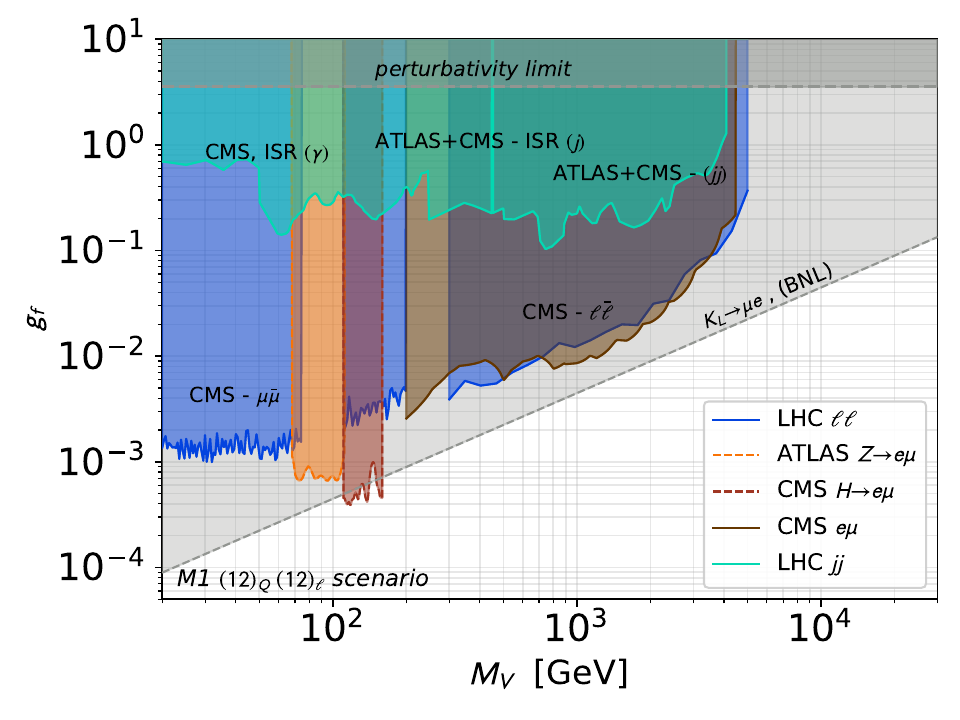}
}%
\hspace{0.02\textwidth}
\subfloat[]{%
\includegraphics[width=0.47\textwidth]{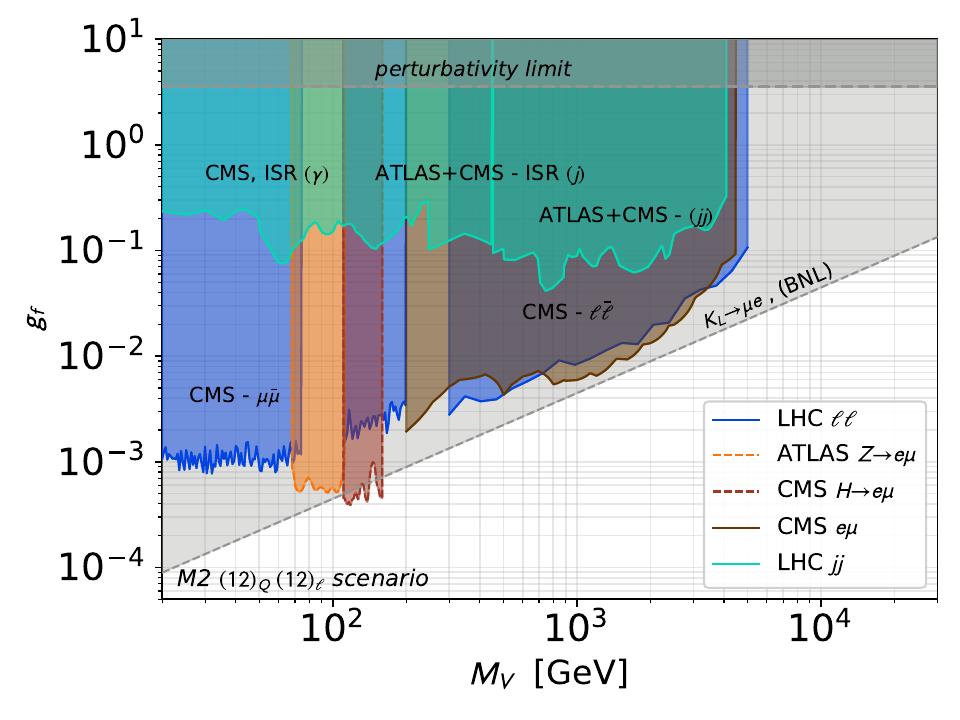}
}%
\caption{LHC limits on our $SU(2)_f$ model with $12_Q$ and $12_\ell$ flavour alignment for (a) scenario (LH) (b) scenario (RH) (c) scenario (M1) and (d) scenario (M2). Aquamarine regions indicate di-jets driven limits~\cite{ATLAS:2019fgd,CMS:2019gwf,ATLAS:2018qto,CMS:2016ltu,ATLAS:2019itm,CMS:2019emo,CMS:2019xai}, the blue regions are for di-muon limits~\cite{CMS:2019buh,CMS:2021ctt} and the brown exclusions from LFV $e\mu$ search~\cite{CMS:2022fsw}. The orange exclusion regions at low mass are recast from LFV $H$ and $Z$ decay searches~\cite{ATLAS:2019old,ATLAS:2021bdj,CMS:2023pte}. We overlaid the limit from pure flavour-transfer processes in dashed grey: $K_L \to \mu e$~\cite{BNL:1998apv}, $CR(\mu \to e)_{Au}$  ~\cite{SINDRUMII:2006dvw} and $K\to \pi \nu_e \nu_\mu$~\cite{NA62:2021zjw}. }
\label{fig:LHC1212}
\end{figure}

Combining the limits derived above, we show in Fig.~\ref{fig:LHC1212} the constraints on the four scenarios for mixing quark and lepton couplings, (LH), (RH), (M1) and (M2) in the simplest case of a flavour alignment  $(12)_Q \ (12)_\ell$. This implies that the flavour non-diagonal bosons $\Wtp, \Wtm$ can be produced via $d \bar s$ or $u \bar c$ initial quark states, then decay into $e \bar \mu$ LFV final states with very small (if any) backgrounds at the LHC. In the large mass limit, the LFV constraints (brown regions) are equivalent to the simpler di-lepton resonance searches driven by the on-shell production of the $\Zt$ boson (blue regions). The latter has a larger background but also stronger production rates since it can be produced directly from $u \bar u$ and $d \bar d$ initial states.

At low masses however, the QED background increases significantly (and the sea quark component of the parton distribution function of the proton dominates), making LFV searches (orange regions) significantly more powerful than the standard di-lepton ones. In Fig.~\ref{fig:LHC1212}, we have put these limits in dashed to indicate that they are obtained from a recasting which makes the assumption of a similar signal efficiency for this particle as for the $Z$ (resp. $H$ boson). Nonetheless, the order of magnitude improvement that these searches provide for our class of NP candidates calls both for a more in-depth re-interpretation, and if possible a dedicated analysis by the experimental collaborations. Similarly, the very low mass regime (below the sideband of the $Z$ boson searches of $60$ GeV) is currently dominated by the CMS search of di-muon resonances, but could most probably see a strong improvement if there were a dedicated LFV search in that region. 

In all the plots of  Fig.~\ref{fig:LHC1212}, we have overlaid in aquamarine the di-jet searches. They are significantly weaker than the lepton-driven ones due to the much larger di-jet background, but apply regardless of the lepton flavour orientation of the model. An interesting feature is a strong reduction in the low mass reach of the (M1) model compared to the others. This is due to the fact that NP in this scenario couples only to down quarks with half the electric charge of up quarks and that these limits are derived by searching for an Initial State Radiation photon.
In all of the scenarios shown in Fig.~\ref{fig:LHC1212}, LHC limits can probe the parameter space close to the best available flavour transfer limit corresponding to the $K_L \to e \mu$ process.  The best search strategies are the ones that rely on flavour-violating leptonic final states. Additionally, as we will see in the next section, a complete numerical fit of the flavour observables in the presence of spurion leads to large parts of the flavour-compatible parameter space in reach of the LHC searches described above.
\begin{figure}[t]
\centering
\subfloat[]{%
\includegraphics[width=0.47\textwidth]{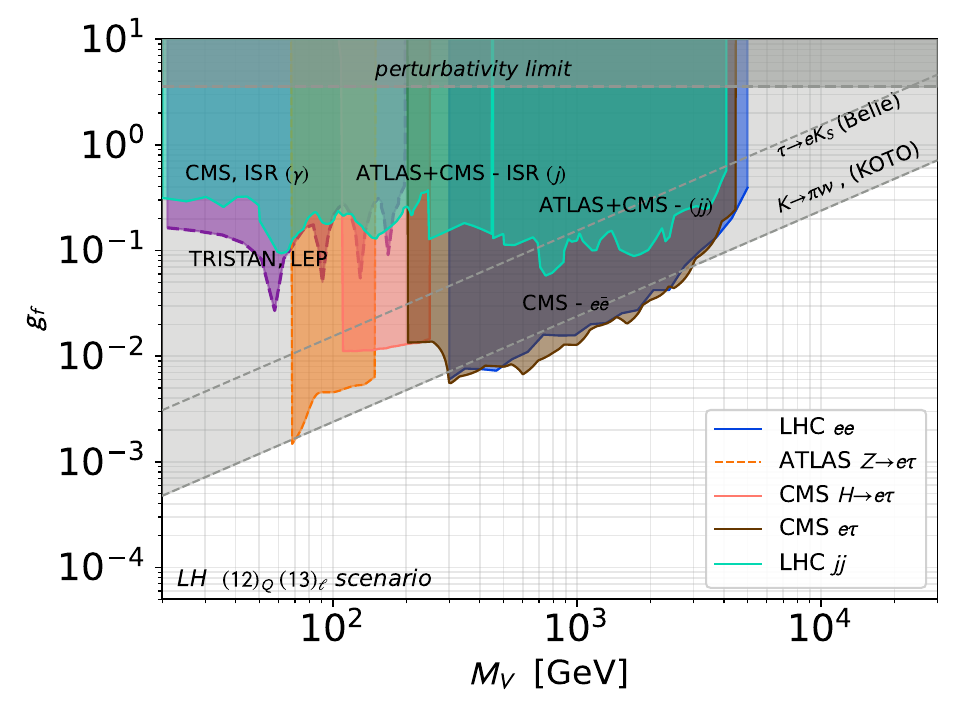}
}%
\hspace{0.02\textwidth}
\subfloat[]{%
\includegraphics[width=0.47\textwidth]{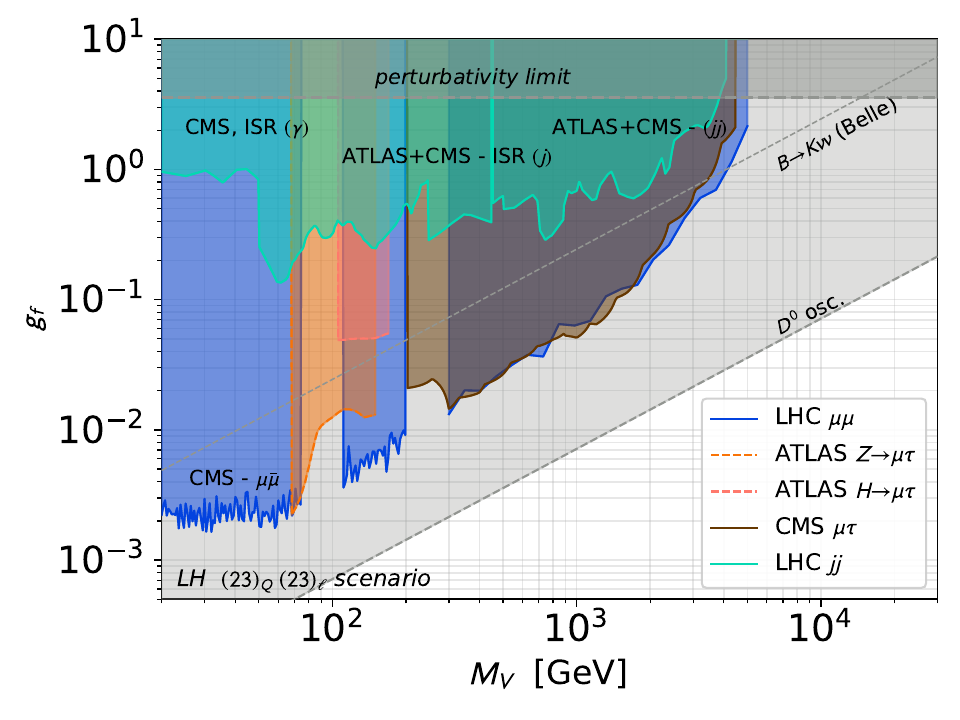}
}%
\caption{LHC limits on our $SU(2)_f$ model for the scenario (LH) with (a) $(12)_Q$ and $(13)_\ell$ flavour alignment  (b) $(23)_Q$ and $(23)_\ell$ flavour alignment. Aquamarine regions indicate di-jets driven limits~\cite{ATLAS:2019fgd,CMS:2019gwf,ATLAS:2018qto,CMS:2016ltu,ATLAS:2019itm,CMS:2019emo,CMS:2019xai}, the blue regions are for di-muon limits~\cite{CMS:2019buh,CMS:2021ctt} and the brown exclusions from LFV $ e\tau$ (a) and $\mu\tau$ searches (b) from~\cite{CMS:2022fsw}. The orange exclusion regions at low mass are recast from LFV $H$ and $Z$ decay searches~\cite{CMS:2021rsq,ATLAS:2022uhq}. The purple exclusion at low masses is from the earlier $e^+e^-$ colliders as obtained in~\cite{Leike:1998wr}.
We overlaid selected limits from pure flavour-transfer processes in dashed grey: $\tau \to e K_S$~\cite{Belle:2010rxj}, $K\to \pi \nu_e \nu_\mu$~\cite{NA62:2021zjw}, $B\to K^* \nu_\tau \nu_\mu$~\cite{Belle:2017oht} and  $D^0$ oscillations~\cite{ParticleDataGroup:2020ssz}. }
\label{fig:LHC_LHflavour}
\end{figure}

In a second time, we consider the possibility of a different flavour alignment, using the (LH) scenario as a reference. We show in Fig.~\ref{fig:LHC_LHflavour} the limits for  $(12)_Q \ (13)_\ell$ and $(23)_Q \ (23)_\ell$ alignments. The $(12)_Q \ (13)_\ell$ has the same production rates as the previous case, but dramatically different final state leptons since it does not couple to muons. In particular, the low mass di-muon constraints from CMS do not apply and are replaced by the old LEP and Tristan constraints~\cite{Leike:1998wr}. LFV searches based on the $e \tau$ final states are recast and give the dominant constraints down to $70$ GeV. While the limits could be in principle recast to lower masses as the experimental results include an overflow, it is clear that our approximation of equal efficiency would not apply there, and a complete re-interpretation study is therefore needed. At higher masses, the di-jet constraints become more competitive with di-electron and LFV $e \tau$ limits. For the $(23)_Q \ (23)_\ell$, the production rates are significantly reduced since the $\suX$ boson does not couple to first-generation quarks anymore. Di-muon searches at low energy become available again however, and LFV $\mu \tau$ limits play an important role at larger masses. In this case, the dominant limits from flavour physics arise from the $D^0$ meson oscillation and will be also somehow weakened once including the possibility of small spurions.

\vspace{0.3cm}
Finally, let us comment on the recent results by the CMS collaboration~\cite{CMS-PAS-HIG-22-002} on the $pp \to e \mu$ processes that have revealed an anomaly at $146$  GeV with a global (local) significance of $2.8\sigma$ ($3.8\sigma$). 
Based on the re-interpretation of the corresponding ATLAS search~\cite{ATLAS:2019old}, we can compare the compatibility of both analyses since they look at the same processes with an equivalent dataset. 

\begin{figure}[t!]
\centering
\hspace{0.02\textwidth}
\subfloat[]{%
\includegraphics[width=0.6\textwidth]{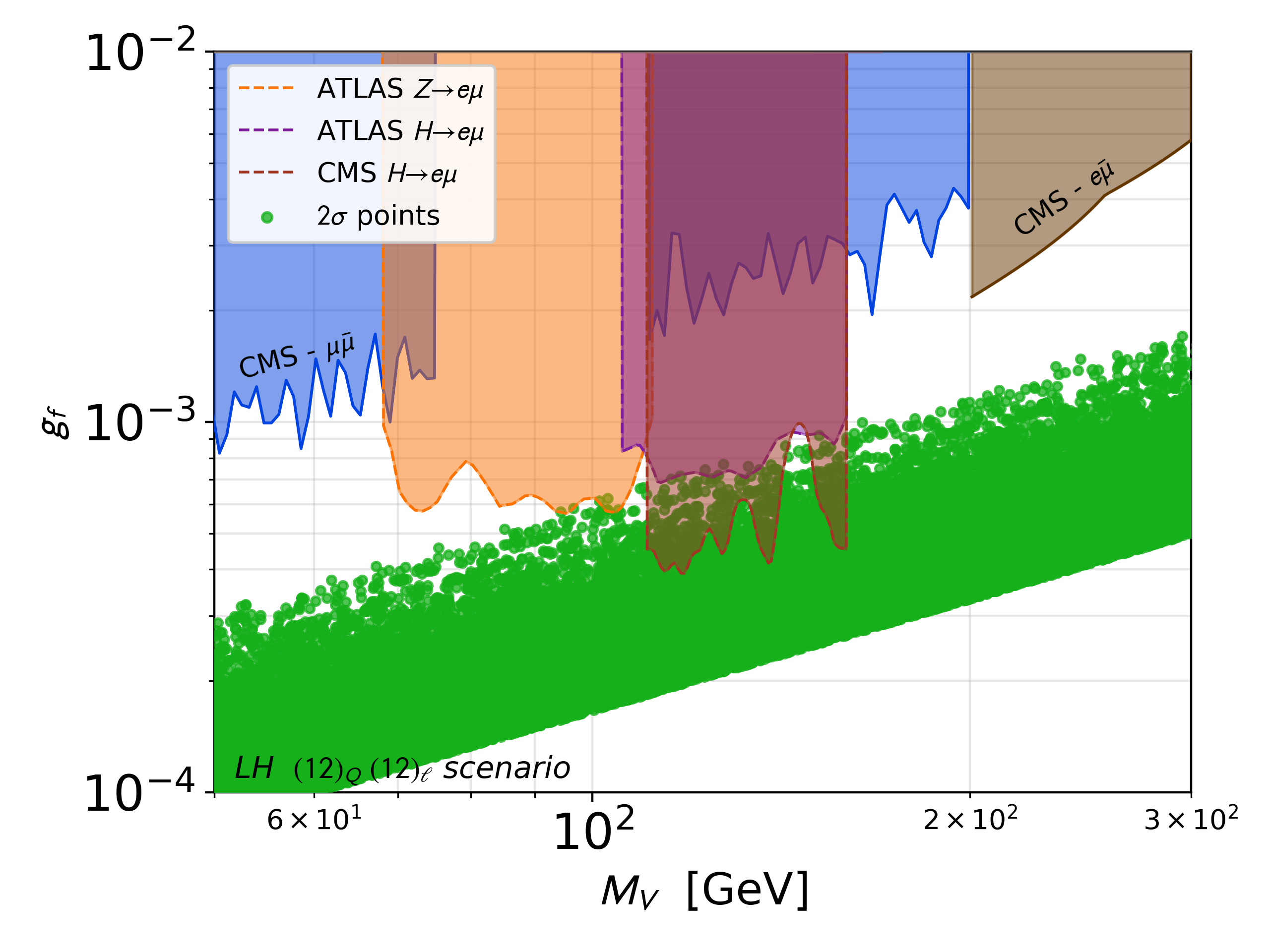}
}%
\caption{ Constraints on the $(12)_Q (12)_\ell$ scenario, with a focus on the small mass region. The green points pass the full flavour fit described in Sec.~\ref{sec:flavourobs} at $2 \sigma$. The blue regions are for di-muon limits~\cite{CMS:2019buh,CMS:2021ctt} and the brown exclusions from LFV $e\mu$ search~\cite{CMS:2022fsw}. The orange, purple and red exclusion regions at low mass are respectively obtained from recasted $Z\to e \mu$ ATLAS search~\cite{ATLAS:2021bdj}, recasted $H\to e \mu$ ATLAS search~\cite{ATLAS:2019old} and a $X\to e \mu$ dedicated NP search at CMS~\cite{CMS:2023pte}.
}
\label{fig:LHC_excess}
\end{figure}

From Fig.~\ref{fig:LHC_excess} the CMS excess is clearly visible as a strong reduction in the experimental sensitivity around $146$ GeV. With the assumption that the signal shape and efficiency for a $146$ GeV signal follow from the ones of the Higgs boson (as described in Sec.~\ref{sec:LHC}). While our analysis of the ATLAS result is limited by the lack of control of the background uncertainties, we recover to a large extent the conclusions of Ref.~\cite{Primulando:2023ugc}: while there is a mild excess in the ATLAS data, it is at a slightly lower mass than the peak of the CMS excess and does not conclusively neither corroborate nor disprove the CMS result.
On the other hand, we note that both analyses are probing currently a completely unexplored region in sensitivity which is additionally fully allowed by flavour observables. As described in Sec.~\ref{sec:LHC}, there is no other channel at the LHC with a similar level of sensitivity in our $\suX$ models and it will be very insightful to follow the evolution
of this excess with the full Run-3 dataset. Finally, we show in  Fig.~\ref{fig:LHC_excess} the results of the full numerical scan that will be described in detail in the next section. We simply point out here that all green points, which are generated only in a range $M_V / g_f < 600 $ TeV, pass the global fit within $1\sigma$ of the SM. Thus flavour constraints in our model are still compatible with the signal seen at CMS.

\section{The case of left-handed $\suX$}
\label{sec:flavourobs}

In this section, we make an in-depth study of the intensity frontier constraints which apply to our $\suX$ scenarios. We consider the (LH) scenario in which both the left-handed quarks $Q_L$ and left-handed leptons $L$ are part of a $\suX$ doublets.
We will both implement a full numerical approach and present for illustration purposes semi-analytical expressions with three different flavour alignments depending on the one relevant for the observable considered:
\begin{itemize}
    \item  $(12)_Q$$(12)_L$: In which the fermion gauge eigenstates are aligned with the first and second generation of mass eigenstates for both left-handed quarks and left-handed leptons. It corresponds to an effective $\Zt$ charge  $\textrm{Diag}(1,-1,0)$.
    \item $(23)_Q$$(23)_L$: Similarly, it corresponds to an effective $\Zt$ charge  $\textrm{Diag}( 0,1,-1)$ for both the quarks and the leptons.
    \item $(12)_Q$$(13)_L$: Corresponding to an effective $\Zt$ charge  $\textrm{Diag}(1,0,-1)$ for the leptons, and $\textrm{Diag}(1,-1,0)$ for the quarks.
\end{itemize}

In all of the semi-analytical expressions, we will assume a ``down'' basis as described in the previous section, and thus use the CKM matrix to obtain the rotation matrix for the left-handed up quarks from the down quarks' one. Since we will be expanding around definite flavour alignment, the expansion parameters will be the rotation angles from the gauge fermion basis to the mass basis (taken to be small). Assuming negligible phases, we will parameterise the rotation matrices $V_x$,  $V_y$ and  $V_z$  from Eq.~\eqref{eq:su2decomp} by the angles $\theta_{f 23}$,  $\theta_{f 13}$ and  $\theta_{f 12}$. As pointed out in Sec.~\ref{sec:gaugesector}, the ordering of the rotation matrices is relevant so that these angles truly represent the mixing between the corresponding fermion generations. For a $(13)$ flavour alignment, we therefore use rotation matrices parameterised as $V_f = V_x V_z V_y$ and for a $(23)$ flavour alignment $V_f =  V_y V_z V_x$. For our full numerical scans we rely on the Eq.~\eqref{eq:su2decomp} with generic angles.

As the angles in the rotation matrices will also parameterise the breaking of the global symmetries of the effective Lagrangian, we will refer to them as ``spurions'' in the following. In a complete UV theory, they would of course arise only as a by-product of the diagonalisation of the mass matrices, and could be expressed in terms of the spurions (for instance Yukawa couplings with a global charge), we give an example of such a model in App.~\ref{sec:appUV}.

All the operators involving quarks will be typically defined in an effective Hamiltonian (thus with a minus sign w.r.t the effective Lagrangian) to follow the standard practice.

\subsection{Mesons oscillations} 

The four measured mass mixing $\Delta M_{ij} $ between neutral mesons, $D_0 - \bar{D}_0$ , $K_0 - \bar{K}_0$ and the $B_s$ and $B_d$ systems, allow us to probe the $uc$, $ds$, $bs$ and $bd$ flavour-violating couplings respectively. Remarkably, such mixings cannot arise solely from a gauge boson exchange in the absence of an additional mass mixing. This is a direct consequence of the global symmetries of the effective Lagrangian underlined in the previous sections. In the limit where all three bosons have the same mass, the partial unitarity of the CKM matrix in the $1-2$ or $2-3$ sectors ensures an additional suppression of these parameters.

There is an extensive literature on the translation between the current dataset and the constraints on effective couplings (see e.g.~\cite{Charles:2020dfl, Aebischer:2020dsw,DiLuzio:2019jyq,King:2019lal,DiLuzio:2017fdq,Jubb:2016mvq}  for some recent results). In the following, we will thus concentrate on exploring directly the expressions for these couplings. For an oscillation $q_i \bar q_j \to \bar q_i q_j $, the relevant operator in our case is
\begin{align}
\mathcal{O}_{1,ij} & = (\bar{q_i} \gamma^\mu P_L q_j) \, (\bar{q_i} \gamma^\mu P_L q_j)\, , 
\end{align}
belonging to the effective Hamiltonian
\begin{equation}
\label{eq:normCi}
{\cal H}_{\rm eff} \supset   \sum_{a=1}^{5} C_{a,ij} (\mu) O_{a,ij}  + h.c\ .
\end{equation}
We leave a more thorough description of this Hamiltonian and of the constraints on meson oscillations to Appendix~\ref{sec:appFV}. For the full numerical results, we will rely on a recent implementation of these observables in \superiso. 

In the limit of $(12)_Q$ alignment, we have for instance (note that the sign is positive since the coefficients are defined from the effective Hamiltonian): 
\begin{align}
\label{eq:spurionOsc}
\mathcal{C}_{1,ij} & = \frac{g_f^2}{8 M_V^2}\times \begin{cases}
         \ \thQsb^2  &\textrm{    for $i,j = b,s $} \\
       \   \thQdb^2   &\textrm{     for $i,j = b,d $} \\
     \    \thQdb^2 \thQsb^2   &\textrm{     for $i,j = s , d $} \\
           \     1.65 \cdot 10^{-3} \, \thQdb^2  + \mathcal{O}(\lamCKM^5)   &\textrm{     for $i,j = c,u $} 
      \end{cases}
\end{align}
Thus, the strongest constraints coming from meson oscillations (see e.g.~\cite{Charles:2020dfl, Aebischer:2020dsw,DiLuzio:2019jyq,King:2019lal,DiLuzio:2017fdq,Jubb:2016mvq}) are typically suppressed by spurions in these constructions. This reflects the fact that they correspond to a $\Delta F = 2$ flavour violation and not to a flavour transfer mechanism as discussed in Sec.~\ref{sec:flavtrans}. The case of the $D_0$-oscillations is however specific. Since we rely on the ``down basis'' for the definition of the spurions, the up quark rotation matrix is fixed by the CKM matrix. The spurion suppression observed in Eq.~\eqref{eq:spurionOsc} is then a consequence of the approximate unitarity of the Cabibbo sub-matrix. If we instead rely on a $(23)_Q$ flavour alignment, we obtain
\begin{align}
\mathcal{C}_{1,cu} & = \frac{g_f^2}{8 M_V^2}\times     \    4.9 \cdot 10^{-2}  + \mathcal{O}(\lamCKM^5)  \ ,
\end{align}
and no spurion suppression, in line with the result of~\cite{Guadagnoli:2018ojc}.

\subsection{Rare meson decays}

The second class of observables that we will consider are rare mesonic decays, which could be mediated by the off-shell exchange of a $\suX$ gauge boson.

\paragraph{Semi-leptonic LFUV meson decays}
We will use the standard normalisation for semi-leptonic transitions $q_i \to q_j \ell \ell$:
\begin{equation}
\label{eq:normC9C10}
{\cal H}_{\rm eff} \supset  -  \frac{4 G_F}{\sqrt{2}} \sum_{i,j \in (b,s), (s,d)} V_{ti}  V^*_{tj} 
                     \sum_{a=1}^{10} C_a (\mu) O_a  + h.c.\ ,
\end{equation}
where $G_F$ is the Fermi constant, and $V_{tj},V_{ti}$ are CKM matrix elements.
The relevant operators in our case are $O_9$ and $O_{10}$, and their primed counterpart for right-handed fields:
\begin{align*}
     O^{sb\ell \ell}_9 &=\frac{e^2}{(4\pi)^2}(\overline{s}_L\gamma_\mu b_L)\overline{\ell}\gamma^\mu \ell\ , \\
     O^{sb\ell \ell}_{10} &=\frac{e^2}{(4\pi)^2}(\overline{s}_L\gamma_\mu b_L)\overline{\ell}\gamma^\mu \gamma_5 \ell\ .
\end{align*}

Semi-leptonic lepton-flavour universality violating (LFUV) observables have seen an intense scrutiny in recent years due to a string of anomalies recorded by the LHCb collaboration. The recent measurement of the semi-leptonic flavour non-universal ratios $R_K$ and $R_{K^*}$ by the LHCb collaboration, with new values compatible with $1$, now points toward a uniform shift for both electron and muon final states~\cite{LHCb:2022qnv,LHCb:2022zom}. We will include the large available experimental dataset~\cite{LHCb:2020lmf,LHCb:2020gog,LHCb:2021xxq,LHCb:2021zwz,LHCb:2021vsc,CMS:2022mgd,BELLE:2019xld,CMS:2018qih} as implemented in \superiso.

Simple semi-analytical results can be obtained however by matching the results from global fits on the effective parameters $C_9$ and $C_{10}$ (see e.g.~\cite{Alguero:2021anc,Altmannshofer:2021qrr,Gubernari:2022hxn,Ciuchini:2021smi,Hurth:2021nsi,Mahmoudi:2022lmp,Ciuchini:2022wbq} for some of the most recent results). 
For our (LH) scenario, we have automatically $C_9= - C_{10}$ and for the $(12)_\ell$ flavour alignment $ C_9^{ee} = -  C_9^{\mu\mu}$ with:
 \begin{align}
     C_9^{sb ee} &= 0.015 \, \thQsb  \, \left( \frac{100 \, \textrm{TeV} }{M_V/g_f} \right)^4  
          \\
     |C_9^{dsee}| &=  0.14 \, \thLemu \,\left( \frac{100 \, \textrm{TeV} }{M_V/g_f} \right)^4  
          \ .
 \end{align}
With this flavour alignment, sizeable effects in the $b \to s$ transitions would thus require a low $M_V / g_f$ around $30$ TeV and large $\thQsb$, while predicting an opposite effect between $\mu \mu$ and $e e$ final states. 

\paragraph{Flavour transferring meson decays $q_i \to q_j \ell_i \bar{\ell}_j$}
Flavour-transferring decays are particularly important in the sense that they are unsuppressed in the spurion approach, since they preserve the global $SU(2)_f$ invariance as long as the emitted lepton flavour doublet  matches the quark ones. 
We will use the same normalisation as for the standard $sd$ and $bs$ operators in Eq.~\eqref{eq:normC9C10} to simplify the comparison with the usual WET operators. 

The first decays we consider are the $K^+ \to \pi^+ \mu e$ processes which are constrained by:
\begin{align}
    &{\textrm{BR}}(K^+ \to \pi^+ \mu^+ e^-) < 1.3 \times 10^{-11} \qquad ({\textrm{E865}} ~\textrm{\cite{Sher:2005sp,Workman:2022ynf}}) \\
    &{\textrm{BR}}(K^+ \to \pi^+ \mu^- e^+) < 6.6 \times 10^{-11} \qquad ({\textrm{NA62}} ~\textrm{\cite{NA62:2021zxl}}) \ ,
\end{align}
with a simple approximation for the new physics prediction for the case of a purely left-handed interaction with $C_9 = -C_{10}$ being given by~\cite{Littenberg:1993qv} (see~\cite{Davidson:2018rqt} for a complete treatment). This corresponds to:
\begin{align}
    \frac{\Gamma(K^+ \to \pi^+ \mu^+ e^-)}{\Gamma(K^+ \to \pi^0 \mu^+ \nu)} = \frac{\alpha_{\rm em}^2}{\pi^2} \left|C^{sde\mu}_9\right|^2 \frac{|V_{td}^*V_{ts}|^2}{|V_{us}|^2} \ ,
\end{align}
and a similar relation for the $\mu^- e^+$ final states by exchanging $C^{sde\mu}$ and $C^{sd\mu e}$. These processes play an important role in constraining $\suX$ gauge interactions for $(12)_Q (12)_\ell$ flavour alignment. For instance, in the case of the (LH) scenario, we obtain
\begin{align}
    \textrm{BR}(K^+ \to \pi^+ \mu^+ e^-) \sim 6.1 \cdot 10^{-12} \times\left( \frac{100 \, \textrm{TeV}}{M_V/g_f}\right)^4 \times  \begin{cases}
       \  1  \textrm{    for $12_\ell$ alignment} \\
      \ \thLmutau^2  \textrm{    for ${13}_\ell$ alignment} \
     \end{cases} \ .
\end{align}
Similar LFV limits can be derived for heavy flavour meson decays, in particular for $B \to K$ and $B \to \pi$ decays, that we describe in Appendix~\ref{sec:appFV}. They are relevant for $(13)_Q$ and $(23)_Q$ flavour alignments in the quark sector.

\paragraph{Fully leptonic meson decays}

Like the semi-leptonic case, these observables are not a priori protected and can be generated by unsuppressed flavour transfer operators.
We focus first on kaonic processes such as $K_L \to \mu e$, which represent an important smoking gun of our models. They are strongly constrained experimentally, with the limit given by
\begin{align}
    \textrm{BR}(K_L \to  \mu^\pm e^\mp) < 4.7 \times 10^{-12} \qquad (\textrm{BNL}~\textrm{\cite{BNL:1998apv}}) \ ,
\end{align}
and the new physics prediction is given for our (LH) scenario by
\begin{align}
     \textrm{BR}(K_L \to  \mu^+ e^-) =  \frac{1}{\Gamma_{K_L}} \frac{M_K f_K^2}{128 \pi^3 } \, \alpha_{\rm em}^2 & G_F^2 |V_{td}^*V_{ts}|^2 \left( 1 - \frac{m_\mu^2}{M_K^2}  \right)^{3/2} \nonumber \\
     & \times \left( |C_9^{sd\mu e} + C_9^{sd e \mu \, *}|^2 + |C_{10}^{sd \mu e } + C_{10}^{sd e\mu \, *}|^2 \right)
\end{align}
with $f_K = 0.155$ GeV~\cite{FlavourLatticeAveragingGroupFLAG:2021npn}. The above expression is in very good agreement with the results obtained with the simpler ratio method in~\cite{Littenberg:1993qv}.
Note that we must sum over both $\mu^+ e^-$ and $\mu^- e^+$ final states to compare with the experimental result, with the later process obtained by exchanging $\mu$ and $e$. Similar expressions for the case of heavier mesons are given in Appendix~\ref{sec:appFV}.

In our benchmark (LH) scenario, this leads to the spurion-independent result
\begin{align}
    \textrm{BR}(K_L \to  \mu^\pm e^\pm) = 1.2 \cdot 10^{-10} \left( \frac{100 \, \textrm{TeV}}{M_V/g_f}\right)^4 \times  \begin{cases}
         \ 1 \quad  \textrm{     for $(12)_\ell$ } \\
      \ \thLmutau^2  \quad  \textrm{    for ${(13)}_\ell$ } \
     \end{cases} \ .
\end{align}
For this last (and dominant constraint), this simplified approach leads to limits on the effective coupling within $\sim 5-15 \%$ of the single-operator results from~\cite{Davidson:2018rqt}.\footnote{Note that the expression for the $P \to \ell_1 \ell_2$ has a missing factor of $4$ compared to, e.g.~\cite{Carpentier:2010ue,Crivellin:2015era,Becirevic:2016zri} that is however included in their numerical results.} This limit constitutes often the dominant constraint on a given parameter point, it is however sensitive to the mixing between both first generation of leptons as $\thLemu^2$ and will be weakened in the full numerical scan.

\paragraph{Invisible meson decays $s\to d \nu \nu $, $b\to s \nu \nu $}

Rare meson decays with neutrinos in the final states are peculiar in that the flavours of the outgoing neutrinos do not impact the final result (apart from the fact that they may not interfere with the SM). Flavour transfer processes can thus proceed regardless of the flavour alignment in the lepton sector.

For neutrino operators, we use the convention:
\begin{equation}
\label{eq:normneutrino}
{\cal H}_{\rm eff} \supset -  \frac{4 G_F}{\sqrt{2}}   V_{ts} V^*_{td}\frac{\alpha_{\rm em}}{4 \pi} \sum_{ij}  C^{dsij}_L (\mu) (\overline{d}_L\gamma_\mu s_L) \ \overline{\nu}_i\gamma^\mu (2 P_L) \nu_j \ .
\end{equation}
Recent experimental results have shed new light on these processes, including the recent results from the NA62 and KOTO collaborations~\cite{NA62:2021zjw}:
\begin{align}
 &{\rm Br}(K^+ \to \pi^+ \nu \bar{\nu}) = 10.6^{+4.1}_{-3.5}
\times 10^{-11}\quad  (\textrm{NA62}~\textrm{\cite{NA62:2021zjw}}), \\
 &{\rm Br}(K_L \to \pi^0 \nu \bar{\nu}) <4.9 \times 10^{-9}\quad (\textrm{KOTO}~\textrm{\cite{KOTO:2020prk}}),
\end{align}
The Standard Model (SM) predictions are obtained from penguin and box diagrams~\cite{DAmbrosio:2022kvb,DAmbrosio:2023irq}:
\begin{align}
    \text{BR}(K^+\to \pi^+\nu \bar{\nu})^{\rm SM}    &= (7.86 \pm 0.61)\times 10^{-11}\,,\\
    \text{BR}(K_L\to \pi^0\nu \bar{\nu})^{\rm SM}    &= (2.68 \pm 0.30) \times 10^{-11}\,.
\end{align}
The NP contribution in our scenarios, assuming that it proceeds mostly via the squared contributions, can be obtained as (see e.g.~\cite{DAmbrosio:2022kvb}):
\begin{align}
  &  \Gamma(K^+ \to \pi^+ \bar{\nu}_i \nu_j) = \frac{\kappa_{K^+} s_W^4}{ 3 \lamCKM^{10}} \, |V_{ts}^*V_{td}|^2 \, |C_{L,sd\nu_i\nu_j}|^2 \  \\
   &     \Gamma(K_L \to \pi^0 \bar{\nu}_i \nu_j) = \frac{\kappa_{K_L} s_W^4}{ 3 \lamCKM^{10}} \, \textrm{Im} (V_{ts}^*V_{td} C_{L,sd\nu_i\nu_j})^2 \ ,
\end{align}
where the $K$ decay coefficients are given by, e.g.~\cite{Brod:2010hi}: $\kappa_{K^+} =0.517 \cdot 10^{-10}$ and $\kappa_{K_L} = 2.231 \cdot 10^{-10}$.
For our full numerical estimate, we rely on the \superiso~implementation, based on~\cite{DAmbrosio:2022kvb}, thus including also the interference with the SM processes for the flavour-preserving neutrino final states. A semi-analytical formula for our example (LH) can be easily obtained as 
\begin{align}
    \textrm{BR}(K^+ \to \pi^+ \nu \nu) \sim 3.6 \cdot 10^{-12} \times\left( \frac{100 \, \textrm{TeV}}{M_V/g_f}\right)^4 \ ,
\end{align}
which is fully spurion-independent thanks to the summation of all outgoing neutrino flavours and thus provides a bound independent of the leptonic flavour structure in this model. In practice, it implies that ratios $M_V /g_f$ lower than $\sim 60$ TeV are not accessible for
(LH) $\suX$ models, as long as the couplings are dominantly to light quarks. Since SM interference may occur, the precise limit will be slightly more parameter-dependent, with the above limit fully relevant once the NP-only
contributions dominate.
Additionally, this constraint relies strongly on having a $\suX$ interaction with neutrinos so that it vanishes for, e.g. the (RH) model described  in Sec.~\ref{sec:SU2theo}.

\vspace{0.2cm}

Similar limits can be obtained for the $B \to K \nu \nu $ processes. The result is typically expressed in terms of the ratio with the SM predictions~\cite{Buras:2014fpa,Crivellin:2015era,Becirevic:2023aov}:
\begin{align}
   R_K^{\nu \bar \nu} ~\equiv~  \frac{{\rm Br}(B^+ \to K^+ \nu \bar{\nu})}{{\rm Br}(B^+ \to K^+ \nu \bar{\nu})_{\rm SM}} = \frac{1}{3} \frac{ \sum_{i,j} | C^{bsij}_L|^2 }{|C^{bs, SM}_L|^2} \ ,
\end{align}
and one can use $C^{bs, SM}_L \simeq -6.32$ and ${\rm Br}(B^+ \to K^+ \nu \bar{\nu})_{\rm SM} = (3.98 \pm 0.47) \times 10^{-6}$ and the numerator include the SM contribution for same-flavour neutrino final states. The same relation also applies for the $B^{0} \to K^* \nu \bar{\nu}$ process, with ${\rm Br}(B^{0} \to K^*  \nu \bar{\nu})_{\rm SM} = (9.19 \pm 0.99) \times 10^{-6}$. On the experimental side, the current limits on these yet undetected processes are:
\begin{align}
    &{\rm Br}(B^+ \to K^+ \nu \bar{\nu}) < 1.6 \times 10^{-5} \,, \  R_K^{\nu \bar \nu}< 4.3 \qquad (\textrm{BaBar} ~\textrm{\cite{BaBar:2013npw}}) \\
    &{\rm Br}(B^0 \to K^* \nu \bar{\nu})< 2.7 \times 10^{-5} \,, \  R_{K^*}^{\nu \bar \nu}< 2.7\qquad (\textrm{Belle} ~\textrm{\cite{Belle:2017oht}}) \ ,
\end{align}
For our example (LH) scenario with $(12)_Q$ flavour alignment, we obtain :
\begin{align}
   R_K^{\nu \bar \nu}  - 1 \sim 1.2  \times\left( \frac{5 \, \textrm{TeV}}{M_V/g_f}\right)^4 \times  \begin{cases}
         \ 0.5 \thQsb^2 +  \thQdb^2    \textrm{    for $(12)_Q$ alignment} \\
       \ 1  \textrm{    for $(23)_Q$ alignment} \ .
     \end{cases}
\end{align}
We note that even in the case of an un-suppressed flavour transfer $b \to s \nu_\tau \bar \nu_\mu $ for  a $(23)_Q ,  (23)_\ell$, the limits are typically weaker than the ones induced by $D^0$ oscillations generated by CKM matrix elements (in the case of a down-type basis).

\subsection{Rare lepton decays} 

We start from a low-energy effective theory of the form:
\begin{align}
\label{eq:EFTlepton}
    \Leff ~\supset &~ \sum_{ij}( F_L^{ij} \bar \ell_i \sigma^{\mu \nu } P_L \ell_j F_{\mu\nu} + F_R^{ij}  \bar \ell_i \sigma^{\mu \nu } P_R \ell_j F_{\mu\nu}  + h.c. )  \nonumber \\[0.5em]
&+ \sum_{k \geqslant i, a \geqslant j} \left( C_{LL}^{aijk} (\bar{\ell}_a \gamma^\mu P_L \ell_i) \, (\bar{\ell}_j \gamma^\mu P_L \ell_k)\, + h.c. \right) \ .  
\end{align}
The operators in the first line are generated at loop level in our model, while the second line corresponds to four-fermion operators generated at tree level. Note that the four-lepton operators with similar chiralities can be ordered as $k \geqslant i$ and $a \geqslant j$ using Fierz identities.\footnote{In fact most operators can be ordered as  $a \geqslant i \geqslant j \geqslant k$, the remaining ones should instead follow $a \geqslant j \geqslant k$ and $i\geqslant k$. This last ordering is relevant to obtain a non-redundant basis, and directly impacts the analytical expressions for, e.g.,  the $\tau^- \to  \mu^- \mu^+  e^-$ process.} The dipole operators are typically not ordered, implying a redundancy that translates into a $1/2$ factor in the effective operator expression below.

The loop-induced interactions of a flavour-violating new boson have been studied for decades and several complete formulae exist in the literature, e.g. in~\cite{Leveille:1977rc,Hisano:1995cp,Langacker:2000ju,Chiang:2011cv,Kriewald:2022erk}. We will focus in this work on the case of new heavy bosons, for which the expressions for most relevant observables can be cast in simple forms. At leading order in $m_\ell^2 / M_V^2$ the dipole coefficients are 
\begin{align}
  F_L^{\alpha \beta} &= - \frac{g_f^2 e }{48 \pi^2 M_V^2} \sum_{i,a} \left[ q_{L,a}^{\alpha i } (q_{L,a}^{\beta i })^* m_\alpha \right]\ ,\\[0.5em]
 F_R^{\alpha \beta} &= - \frac{g_f^2 e }{48 \pi^2 M_V^2} \sum_{i,a} \left[ q_{L,a}^{\alpha i } (q_{L,a}^{\beta i })^* m_\beta  \right] \ .
\end{align}

We note that even at the one-loop level, flavour-violating contributions are typically suppressed by spurions insertion. For a flavour orientation $1-2$ we get for instance
\begin{align}
\label{eq:sumbosons}
 \sum_{i,a} q_{L,a}^{i \alpha} (q_{L,a}^{\beta i })^*   &\simeq \frac{3}{4} \begin{pmatrix}
     1&0&-\epsilon_{13}\\
     0&1&-\epsilon_{23}\\
      -\epsilon_{13}&-\epsilon_{23}&0\\
 \end{pmatrix} \ .
\end{align}

\paragraph{Radiative lepton decays}

We can leverage the LFV dipole operators from Eq.~\eqref{eq:EFTlepton} to study the radiative $\mu \to e\gamma$,  $\tau \to e\gamma$ and  $\tau \to \mu \gamma$ decays. The current experimental constraints are
\begin{align}
    \textrm{BR} (\mu \to e \gamma) < 4.2 \cdot 10^{-13} \quad (\textrm{MEG}~\textrm{\cite{MEG:2016leq}}) \\
        \textrm{BR} (\tau \to \ell\gamma) < 3.3 \cdot 10^{-8} \quad (\textrm{BaBar}~\textrm{\cite{BaBar:2009hkt}}) \ .
\end{align}
The branching ratios can be obtained simply from  \cite{Chiang:2011cv,Kriewald:2022erk,Davidson:2022nnl,Buras:2021btx}:
\begin{align}
    \textrm{BR} (\ell_\alpha \to \ell_\beta \gamma) = \frac{(m_\alpha^2 - m_\beta^2)^3 }{4 \pi m_\alpha^3 \Gamma_\alpha} \left( |F_L|^2 + |F_R|^2 \right) \ .
\end{align}
In the case of the (LH) scenario with $12$ orientated leptonic couplings, we thus have
\begin{align}
    \textrm{BR} (\ell \to \ell^\prime \gamma) &\simeq  10^{-11} \times \left( \frac{10 \, \textrm{TeV}}{M_V/g_f}\right)^4 
     \times \begin{cases}
8.0 \, \thLmutau^2 \thLetau^2 \quad (\mu \to e \gamma) \\
1.4 \, \thLetau^2 \qquad\quad (\tau \to e \gamma) \\
1.4\,  \thLmutau^2 \qquad \quad (\tau \to \mu \gamma) 
    \end{cases} \ .
\end{align}
As expected, the $\mu \to e \gamma$ process leads to the strongest constraint but depends on spurions at the fourth power, implying it can be very strongly suppressed in models with  leptonic masses ``aligned'' with gauge group representations.
Interestingly, for the $(13)_\ell$ alignment we get
\begin{align}
    \textrm{BR} (\ell \to \ell^\prime \gamma) &\simeq  10^{-11} \times \left( \frac{10 \, \textrm{TeV}}{M_V/g_f}\right)^4 
     \times \begin{cases}
8.0 \, \thLemu^2 \qquad \quad (\mu \to e \gamma) \\
1.4 \, \thLmutau^2  \thLemu^2  \qquad \quad (\tau \to e \gamma) \\
1.4\,  \thLmutau^2  \quad (\tau \to \mu \gamma) 
    \end{cases} \ ,
\end{align}
where the $\mu \to e \gamma$ process is instead suppressed only at the $\thLemu^2$ level.

Finally, we note that the same dipole operators can be used to generate anomalous lepton magnetic moments. The strong constraints from the flavour transfer observables in our example (LH) scenario will however exclude the parameter space with any visible effect. 

\paragraph{Rare three-body lepton decays}
The second class of flavour-violating decays covers the charged $\mu \to e \bar e e$ and $\tau \to \ell \bar \ell \ell$ processes, which can arise at tree level in our models. The relevant processes are 
\begin{align}
    &\textrm{BR} (\mu \to e \bar e e) < 1.0 \cdot 10^{-12} &&(\textrm{SINDRUM}~\textrm{\cite{SINDRUM:1987nra}}) \nonumber \\
        &\textrm{BR} (\tau \to 3 e) < 3.3 \cdot 10^{-8} \ ,  \textrm{BR} (\tau \to 3 \mu) < 2.1 \cdot 10^{-8}  &&(\textrm{Belle}~\textrm{\cite{Hayasaka:2010np}}) \\
        &\textrm{BR} (\tau^- \to \mu^- e^+ e^-) < 1.8 \cdot 10^{-8}\, , \  \textrm{BR} (\tau \to e^- \mu^+ \mu^-) < 2.7 \cdot 10^{-8}  &&(\textrm{Belle}~\textrm{\cite{Hayasaka:2010np}}) \ , \nonumber 
\end{align}
with other related $\tau$ decays probed by Belle~\cite{Hayasaka:2010np} at the same order.

We use four-fermion effective operators as defined in Eq.~\eqref{eq:EFTlepton}.
In particular, we are interested in all flavour-violating $\ell_a \to \ell_i \ell_j \bar{\ell}_k $ decays. Since the four-fermion operators mediating these decays appear at tree-level in our model, we will focus on their contributions (see e.g.~\cite{Kriewald:2022erk,Davidson:2022nnl} for the full expressions including the dipole and anapole operators).
Following~\cite{Chiang:2011cv,Crivellin:2013hpa,Buras:2021btx}, the branching ratio is given at tree-level by
\begin{align}
    \textrm{BR} (\ell_\alpha \to \ell_i \ell_j \bar{\ell}_k  ) =  \frac{m_\alpha^5}{1536 \pi^3 \Gamma_\alpha} \, \times \begin{cases}
2 |C^{i \alpha i k}_{LL}|^2  \textrm{  if $i=j$ }\\
 |C^{i \alpha j k}_{LL}|^2 \textrm{  if $i\neq j$ }
    \end{cases} \ .
\end{align}
In our (LH) example, we obtain for the $12_\ell$ alignment:
\begin{align}
   \textrm{BR} (\ell_\alpha \to \ell_i \ell_j \bar{\ell}_k)  &\simeq 10^{-9} \times \left( \frac{10 \, \textrm{TeV}}{M_V/g_f}\right)^4  \times \begin{cases}
11.6 \, \thLmutau^2 \thLetau^2 \quad (\mu \to e \bar e e) \\
1.0 \, \thLetau^2 \qquad\quad (\tau \to e \bar \mu \mu) \\
1.0 \,  \thLmutau^2 \qquad \quad (\tau \to \mu \bar e e) 
    \end{cases} \ ,
\end{align}
while for the $13_\ell$ case we get (up to four power of a spurions insertions):
\begin{align}
   \textrm{BR} (\ell_\alpha \to \ell_i \ell_j \bar{\ell}_k)  &\simeq 10^{-9} \times \left( \frac{10 \, \textrm{TeV}}{M_V/g_f}\right)^4 \times \begin{cases}
11.6 \thLemu^2  \qquad\quad (\mu \to e \bar e e) \\
1.0 \, \thLmutau^2 \thLemu^2 \quad (\tau \to e \bar \mu \mu) \\
1.0 \,  \thLmutau^2  \qquad \quad (\tau \to \mu \bar e e) 
    \end{cases} \ .
\end{align}

As expected, these processes lead to constraints significantly more stringent than the loop-induced radiative decays. Saturating both $\tau \to \mu \bar e e$ and $\tau \to e \bar \mu \mu$ limits implies that the $ \epsilon_{13}^2 \epsilon_{23}^2$ suppression of the $\mu \to e \bar e e$ is not enough to escape the limit from SINDRUM, indicating that we either expect a hierarchy between both spurions, e.g. $\epsilon_{13} \ll \epsilon_{23}$, or $\mu \to e \bar e e$ will constitute the strongest experimental limit.
 
\paragraph{Flavour transfer processes: hadronic $\tau$ decays}

While the flavour transfer decays $\ell \to \ell^\prime \bar \nu_\ell \nu_{\ell^\prime}$ do not have any spurion suppression, they are also indistinguishable from the EW-induced decays and will therefore suffer from
a very large SM background. Relevant flavour transfer processes hence typically imply quarks in the final states. The only kinematically-allowed processes are based on $\tau$ decays and require a different alignment between the quark and the lepton sectors (for instance $(12)_Q$ and $(23)_\ell$ lead to $\tau \to \mu K $). In general, for a vector meson $V$ with quark constituent $d_i$ and $d_j$ (in practice, either $s$ or $d$), the processes $\tau \to \ell V$ are given by (using the $C_9$, $C_{10}$ normalisation convention for the semi-leptonic operators:
\begin{equation}
    \textrm{BR} (\tau \to \ell_i V ) = \frac{G_F^2 |V_{t d_i}^*V_{td_j}|^2 \alpha_{\rm em}^2 }{32\pi^3} \frac{f_V^2 m_\tau^3}{ \Gamma_\tau} (1- x_V)\, (-1 +\frac{x_V}{2}- \frac{x_V^2}{2})
    \, \times \left| C^{q_iq_j \ell \tau}_9 \right|^2  
    \ ,
\end{equation}
where we used $x_V = M_V^2 / m_\tau^2$, and the decay constants can be found in, e.g.~\cite{Chang:2018aut,Chen:2020qma}, with $f_\phi = 241 \pm 9 \pm 2 \, \rm MeV$, $f_{K^*} = 204 \pm 7 \, \rm MeV$,  $f_{\rho} = 210 \pm 4 \, \rm MeV$.\footnote{In the case of a $\rho / \omega$ meson final state, the interpolating quark current $(u \bar u \pm d \bar d)/\sqrt{2} $ means that the $C_{9,10}$ coefficients must be replaced by the appropriate combinations.}
The limits on these decays are given by~\cite{Belle:2011ogy}, we quote here some of the most relevant ones:
\begin{align}
    &\textrm{BR} (\tau^- \to  \mu^- \bar{K}^{*0}) < 7.0 \cdot 10^{-8}  \nonumber \\
    &\textrm{BR} (\tau^- \to  e^- \bar{K}^{*0}) < 3.2 \cdot 10^{-8}   \\
&\textrm{BR} (\tau^- \to  \mu^- \phi) < 8.4 \cdot 10^{-8}  \nonumber \ .
\end{align}
Similarly, for the case of a pseudo-scalar meson we get (see e.g.~\cite{Celis:2014asa}):
\begin{align}
    \textrm{BR} (\tau \to \ell_i P ) &= \frac{G_F^2 |V_{t d_i}^*V_{td_j}|^2 \alpha_{\rm em}^2 }{64\pi^3} \frac{f_P^2 m_\tau^3}{\Gamma_\tau} (1- x_P^2)^2\, \times \left| C^{P \ell \tau}_9 \right|^2  \ ,
\end{align}
where for the $\pi^0$ and $K_S$ cases  $C^{\pi^0 \ell \tau} = (C^{uu \ell \tau} - C^{dd \ell \tau})/\sqrt{2}$ and $C^{K_S \ell \tau} = (C^{sd \ell \tau} + C^{ds \ell \tau})/\sqrt{2}$, and $f_\pi = 0.130$ GeV, $f_K = 0.155$ GeV. Existing limits are of similar magnitude as for the vector case, for instance~\cite{Belle:2010rxj}:
\begin{align}
    &\textrm{BR} (\tau^- \to  e^- K_S) < 2.6 \cdot 10^{-8}   \qquad (\textrm{Belle}) \ .
\end{align}
The $K^*$ and $K_S$ decays are particularly interesting in that they are pure flavour transfer processes in the cases of $(12)_Q (23)_\ell$ and $(12)_Q (13)_\ell$ flavour alignments. As an example, for the $(12)_Q (23)_\ell$ alignment, we obtain:
\begin{align}
   \textrm{BR} (\tau^- \to  e^- K_S)   &\simeq 1.8 \cdot 10^{-9} \times \left( \frac{10 \, \textrm{TeV}}{M_V/g_f}\right)^4 \ .
\end{align}

\subsection{Other constraints}

\paragraph{Muon conversion in nuclei} 

Experiments studying the decays of muons typically require a high-luminosity flux which can be captured in a target to study the process $\mu \textrm{A} \to e \textrm{A}$, where $\textrm{A}$ can be any nuclei. One of the most stringent limits has been obtained using gold nuclei by the SINDRUM-II collaboration:
\begin{align}
\label{eq:expCR}
  \frac{\Gamma(\mu^- \, \textrm{Au} \to e^- \, \textrm{Au})}{\Gamma_{\textrm{capt}}(\mu^- \,\textrm{Au})}  ~\equiv~ \textrm{CR} (\mu^- \, \textrm{Au} \to e^- \,\textrm{Au}) < 7 \cdot 10^{-13} \quad (\textrm{SINDRUM-II}~\textrm{\cite{SINDRUMII:2006dvw}}) \ ,
\end{align}
where $\textrm{CR}$ is the conversion ratio for negatively charged muons in the corresponding nuclei.
The dominant contributions to this process arise from mixed lepton/quark operators:
\begin{align}
   \Leff ~\supset~ C_{L}^{e\mu uu} (\bar{e} \gamma^\mu P_L \mu) \, (\bar{u} \gamma^\mu P_L u)\, + C_{LR}^{e\mu uu} (\bar{e} \gamma^\mu P_L \mu) \, (\bar{u} \gamma^\mu P_R u)\,\, + (L \leftrightarrow R \ , u \leftrightarrow d) +  h.c. \nonumber 
\end{align}
Only the vector combination of the quark currents impacts the CR, which is given as~\cite{Kitano:2002mt}:
\begin{align}
    \textrm{CR} (\mu \, \textrm{A} \to e \,\textrm{A}) = \frac{4 m_\mu^5}{\Gamma^A_{\textrm{capt}}} \left| \frac{1}{4 m_\mu} F_R^* D_A + \tilde{C}^{(p)}_{LV} V^{(p)}  + \tilde{C}^{(n)}_{LV} V^{(n)}\right|^2 +  \left| (L \leftrightarrow R ) \right|^2 \ ,
\end{align}
where we have used the normalisation of the dipole operators defined in Eq.~\eqref{eq:EFTlepton}. $V_A^{(n), (p)}$ and $D_A$  denote the nucleus-dependent overlap integrals defined in Ref.~\cite{Kitano:2002mt} (we used the first set of values they provide -- method 1 -- and estimated the uncertainties using the second set). $\Gamma^A_{\textrm{capt}}$ are the total capture rates~\cite{Suzuki:1987jf},  and the effective proton and neutron couplings are defined for our (LH) scenario as:
\begin{align}
    \tilde{C}^{(p)}_{LV}  =  C_{L}^{e\mu uu}  + \frac{1}{2}   C_{L}^{e\mu dd} \ ,   \\
    \tilde{C}^{(n)}_{LV}  = \frac{1}{2} C_{L}^{e\mu uu}  +    C_{L}^{e\mu dd} \ ,
\end{align}
with the coefficients simply reflecting the nucleon quark content. Following, e.g.~\cite{Davidson:2022nnl}, we will consider both the case of $\textrm{Au}$ nuclei~\cite{SINDRUMII:2006dvw} and $\textrm{Al}$ nuclei as they are expected to be studied at MU2e~\cite{Mu2e:2022ggl}. The required nuclear physics inputs are given by~\cite{Kitano:2002mt, Suzuki:1987jf}:
\begin{align}
    (V_{A}^{(p)},V_{A}^{(n)},D_{A}) = \begin{cases}
       \ (0.0974,\, 0.146,\,0.189)  \quad \ \ (\textrm{Au})\\
       \ (0.0161,\, 0.0173,\,0.0362)\quad (\textrm{Al})\\
    \end{cases} , \
    \Gamma^A_{\textrm{capt}} = 10^{6} s^{-1} \times \begin{cases}
       \ 13.07  \quad \  (\textrm{Au})\\
     \ 0.7054 \quad (\textrm{Al})\\
    \end{cases} \nonumber
\end{align}

As an example, in the case of the (L) scenario with $1-2$ oriented couplings we obtain the spurion-independent results (this is a consequence of the CKM requirement we fixed for the up-quark rotation matrices):
\begin{align}
   \textrm{CR} (\mu \, \textrm{Au} \to e \,\textrm{Au}) \sim 2.2 \cdot 10^{-11} \left( \frac{100 \, \textrm{TeV}}{M_V/g_f}\right)^4 \times ( 1 + 18.5 \, \thLemu)\ .
\end{align}
Similarly to the mixed lepton/quark processes in Kaon decays, there is a non-suppressed term that will contribute in the absence of spurion. Quite remarkably, the contribution of both neutron and proton terms leads to the possibility of suppressing these observables for a small rotation angle $\thLemu$. Given the stringent limit shown in Eq.~\eqref{eq:expCR}, we do not expect a strong effect without a significant amount of tuning.

In the $13_\ell$ orientation for the leptonic sector, this process instead appears suppressed as expected:
\begin{align}
\label{eq:CRmuToeInAu}
   \textrm{CR} (\mu \, \textrm{Au} \to e \, \textrm{Au}) \sim 20.0 \cdot 10^{-11} \left( \frac{100 \, \textrm{TeV}}{M_V/g_f}\right)^4 \times |\thLmutau \, \thLemu - 2.3\, \thLemu^2 -0.11\, \thLmutau^2 | \ .
\end{align}
The strong experimental bounds will nonetheless lead to significant constraints on $\thLmutau$ and $\thLetau$, particularly at low $M_V/g_f$.

\paragraph{Muonium oscillations}

Muonium oscillations are $\Delta F = 2 $ processes corresponding to the transition $\mu^- e^+ \leftrightarrow \mu^+ e^- $  . While these measurements usually lead to important constraints for models with light new mediators, they can also probe effective operators (see~\cite{Conlin:2020veq} and the review in~\cite{Davidson:2022jai}). We use a left-right basis for the effective operators:
\begin{align}
    \mathcal{L} ~\supset~ C_{1}^{\mu e\mu e} (\bar{\mu} \gamma^\mu P_L e) \, (\bar{\mu} \gamma^\mu P_L e) + C_{2}^{\mu e\mu e} (\bar{\mu} \gamma^\mu P_R e) \, (\bar{\mu} \gamma^\mu P_R e)  + C_{3}^{\mu e\mu e}  (\bar{\mu} \gamma^\mu P_L e) \, (\bar{\mu} \gamma^\mu P_R e)   +  h.c. \nonumber \ .
\end{align}
Since this process maximally breaks the flavour transfer structure, the effective coupling  is suppressed by four powers of spurions:
\begin{align}
    C_{1}^{\mu e\mu e} =  1.4 \cdot 10^{-8}  \left( \frac{3 \, \textrm{TeV}}{M_V/g_f}\right)^2 \thLetau^2  \thLmutau^2\,  \rm{ GeV}^{-2}\ .
\end{align}
The experimental limit on muonium oscillation~\cite{Willmann:1998gd} corresponds to a bound on the above effective coefficient given by~\cite{Conlin:2020veq}:
\begin{align}
    C_{1}^{\mu e\mu e} <  3.4 \cdot 10^{-8}  \, \rm{ GeV}^{-2} \ .
\end{align}
Thus, we do not expect muonium oscillations to lead to a sizeable constraint on our scenarios.

\subsection{Summary of constraints}

We have shown in the previous sections that the only sets of effective operators which are not suppressed in the limit of definite flavour alignment (and therefore which proceed via flavour transfer) are  the mixed lepton/quark operators. We illustrate this pattern and summarise the large corpus of constraints considered in this section in Table~\ref{tab:summaryintensity}. 
\begin{table}
\begin{center}
	\resizebox{1.\textwidth}{!}{\begin{minipage}{1.11\textwidth}
\begin{tabular}{l l|ccc}
\hline
& &\multicolumn{3}{c}{$SU(2)_f$ flavour alignment}\\
  \rule{0pt}{1.25em} Constraints & Refs. & $(12)_Q$$(12)_\ell$ & $(23)_Q$$(23)_\ell$  & $(12)_Q$$(13)_\ell$  \\[0.5em]
\hline \rule{0pt}{1.25em}$B \to K ee \ (C_9)$ &/& $-\thQsb$ & $+\thLemu \thLetau$ & $-\thQsb$\\
$B \to K \mu\mu \ (C_9)$  &/& $+\thQsb$&$-\thLmutau$&$0$\\ 
$K \to \pi ee \ (C_9)$  &/& $+\thLemu$&$0$&$+ \thLetau$\\
$K \to \pi \mu\mu \ (C_9)$ &/& $-\thLemu$&$+\thQds$ & $\thLemu \thLmutau$\\
\hline\rule{0pt}{1.25em}$\textrm{BR}^{\textrm{ (E865)}}_{K^+ \to \pi^+ \mu^+ e^-} < 1.3 \times 10^{-11}$  & \cite{Sher:2005sp,Workman:2022ynf}&  \cellcolor{blue!25}$1$ & $0$ & $\thLmutau^2$\\[0.4em]
$\textrm{BR}^{\textrm{ (E865)}}_{K^+ \to \pi^+ \mu^- e^+} < 6.6 \times 10^{-11}$  & \cite{Sher:2005sp,Workman:2022ynf}& $0$ & $0$ & $0$ \\[0.4em]
${\rm Br}^{\textrm{ (NA62)}}_{K^+ \to \pi^+ \nu \bar{\nu}} = 1.06^{+0.41}_{-0.35} \times 10^{-10}$  & \cite{NA62:2021zjw}&\cellcolor{blue!25} $1$ & $ \thQds^2$&\cellcolor{blue!25} $1$\\[0.4em]
 $\textrm{BR}^{\textrm{ (BNL)}}_{K_L \to  \mu^+ e^-} < 4.7 \times 10^{-12}$ &\cite{BNL:1998apv}& \cellcolor{blue!25} $1$ & $0$ & $\thLmutau^2$\\[0.4em]
  \hline
  BR$^{\textrm{ (BaBar)}}_{B^+ \to K^+ \nu \nu} <  1.6 \times 10^{-5}$ &\cite{BaBar:2013npw}& $2\thQdb^2 + \thQsb^2 $ & \cellcolor{blue!25} $1$ & $2\thQdb^2 + \thQsb^2 $\\[0.4em]
  BR$^{\textrm{ (LHCb)}}_{B^+ \to K^+ e^-\mu^+}<  6.4 \times 10^{-9}$ &\cite{LHCb:2019bix}& $\thQdb^2$& $\thLetau^2$ & $0$\\[0.4em]
  BR$^{\textrm{ (BaBar)}}_{B^+ \to K^+ \mu^- \tau^+ } < 2.8  \times 10^{-5}$ &\cite{BaBar:2012azg}&$0$ &  \cellcolor{blue!25} $1$  & $0$ \\[0.4em]
 \hline
 \rule{0pt}{1.25em}$K$ oscillations ($C_1$) &\cite{HeavyFlavorAveragingGroup:2022wzx}& $0$ & $\thQds^2$ & $0$ \\
  $D$ oscillations ($C_1$) &\cite{HeavyFlavorAveragingGroup:2022wzx}& $ \thQdb^2$ & \cellcolor{blue!25} $1 - 8 \thQds$& $\thQdb^2$\\
$B_d$ oscillations ($C_1$)&\cite{HeavyFlavorAveragingGroup:2022wzx}&$ \thQdb^2$ & $ \thQdb^2$ &  $\thQdb^2$\\
$B_s$ oscillations ($C_1$)&\cite{HeavyFlavorAveragingGroup:2022wzx}&$ \thQsb^2$ &  $0$ &  $\thQsb^2$\\[0.4em]
 \hline
 \rule{0pt}{1.25em}$\textrm{BR}^{\textrm{ (SINDRUM)}}_{\mu \to e \bar e e} < 1.0 \cdot 10^{-12}$ &\cite{SINDRUM:1987nra}& $0$& $0$ & $\thLmutau^2$\\[0.4em]
  $\textrm{BR}^{\textrm{ (BELLE)}}_{\tau \to 3 \mu} < 2.1 \cdot 10^{-8} $  & \cite{Hayasaka:2010np}& $\thLmutau^2$& $0$ & $0$\\[0.4em]
  $\textrm{BR}^{\textrm{ (BELLE)}}_{\tau \to 3 e} < 3.3 \cdot 10^{-8}$   & \cite{Hayasaka:2010np}& $\thLetau^2$& $0$ & $0$\\[0.4em]
 $\textrm{BR}^{\textrm{ (MEG)}}_{\mu \to e \gamma} < 4.2 \cdot 10^{-13} $  & \cite{Kriewald:2022erk,MEG:2016leq}&  $0$& $\thLemu^2$ &  $\thLetau^2$ \\[0.4em]
 $\textrm{BR}^{\textrm{ (Belle)}}_{\tau \to  e \bar{K}^*} < 3.2 \cdot 10^{-8} $  & \cite{Belle:2011ogy} &  $0$& $0$ & \cellcolor{blue!25} $1$ \\[0.4em]
  $\textrm{BR}^{\textrm{ (Belle)}}_{\tau \to \mu \bar{K}^*} < 7.0 \cdot 10^{-8} $  & \cite{Belle:2011ogy} &  $\thLetau^2$& $\thQdb^2$ & $\thLemu^2$ \\[0.4em]
 \hline
 \rule{0pt}{1.25em}$\textrm{CR}^{\textrm{ (SINDRUM-II)}}_{Au, \mu  \to e } < 7 \cdot 10^{-13} $  &\cite{Kitano:2002mt,Davidson:2022nnl,SINDRUMII:2006dvw}&\cellcolor{blue!25} $1+20\thLemu$& $\thLemu^2$ & $\thLemu (2.3 \thLemu-\thLmutau)$\\[0.4em]
$\mu \bar{e} \to e \bar{\mu} $ oscillations ($C_1$) & \cite{Willmann:1998gd} & $0$& $\thLemu^2$ & $\thLemu^2$ \\[0.7em]
\hline
\end{tabular}
		\end{minipage}}
\caption{Selection of some of the intensity frontier observables used as constraints in this work, underlying the  typical spurion dependence as a function of the dominant flavour alignment of the $\suX$ doublets. We retain the small spurionic rotation angles at order $\theta^2$.}
\label{tab:summaryintensity}
	\end{center}
\end{table}
The typical spurions scaling for the (LH) model and for various choices of flavour orientations is also given in this table.
As expected, observables based on mixed quark/lepton operators provide in most cases the dominant limit on the couplings and masses of the $\suX$ gauge boson. In particular, the limits on $K_L  \to e \mu$  from the BNL experiment dominate for the case $(12)_Q$$(12)_L$, despite being based on two decades-old experimental results. Muon capture and conversion in nuclei also provide a strong constraint since they rely on both $(\bar u u ) (\bar \mu e)$ and $(\bar d d ) (\bar \mu e)$ operators. There again, the current limits are only based on the SINDRUM-II results from two decades ago. Dramatic improvements with up to five orders of magnitude gains on the $\textrm{CR}_{\textrm{Au}, \mu  \to e }$ sensitivities are expected in the COMET~\cite{Moritsu:2022lem} and Mu2e~\cite{Bernstein:2019fyh} experiments. For the simplest $(12)_Q$$(12)_L$ scenario, this would correspond to PeV-scale limits on $M_V / g_f$. 

\subsection{Numerical results}

In order to obtain the full numerical results beyond the approximation presented above, we have implemented -- when not already available -- the relevant observables listed above in the code \superiso. The scanning process relies on the \bsmart~\cite{Goodsell:2023iac} code, using the \multinest~\cite{Feroz:2008xx} routines to converge on phenomenologically relevant regions (as determined by the $\chi^2$ obtained as output of \superiso).
Our input parameters follow the parameterisation described in Eqs.~\eqref{eq:gaugefermionMass},~\eqref{eq:su2decomp} and~\eqref{eq:effcharge}. More details on the scan procedure are given in Appendix~\ref{sec:appScan}.

\begin{figure}[t]
\centering
\subfloat[]{%
\includegraphics[width=0.47\textwidth]{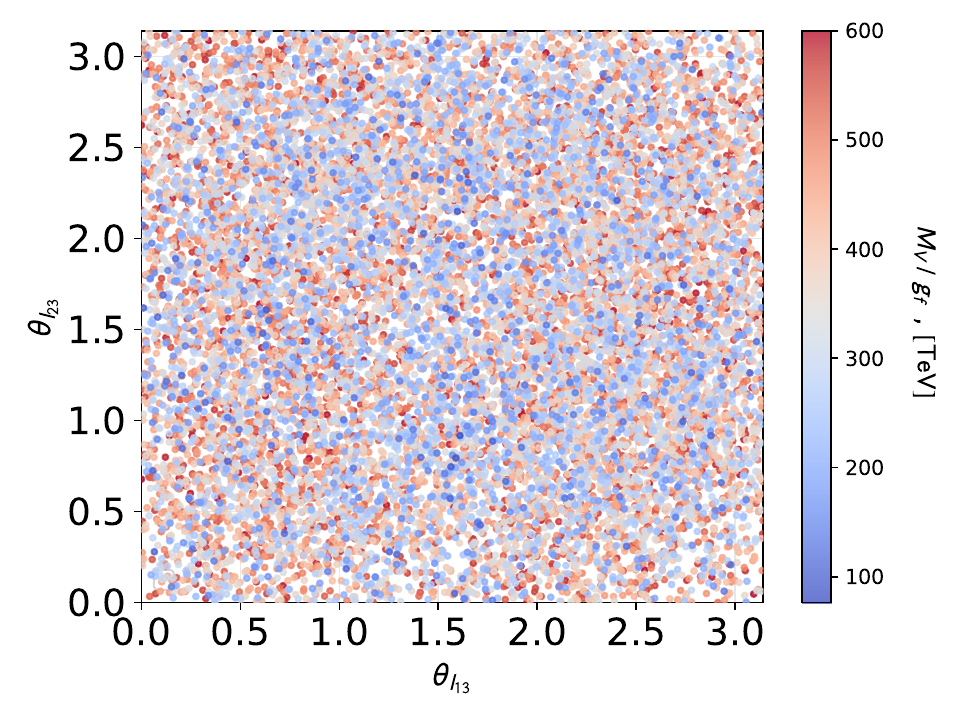}
}%
\hspace{0.02\textwidth}
\subfloat[]{%
\includegraphics[width=0.47\textwidth]{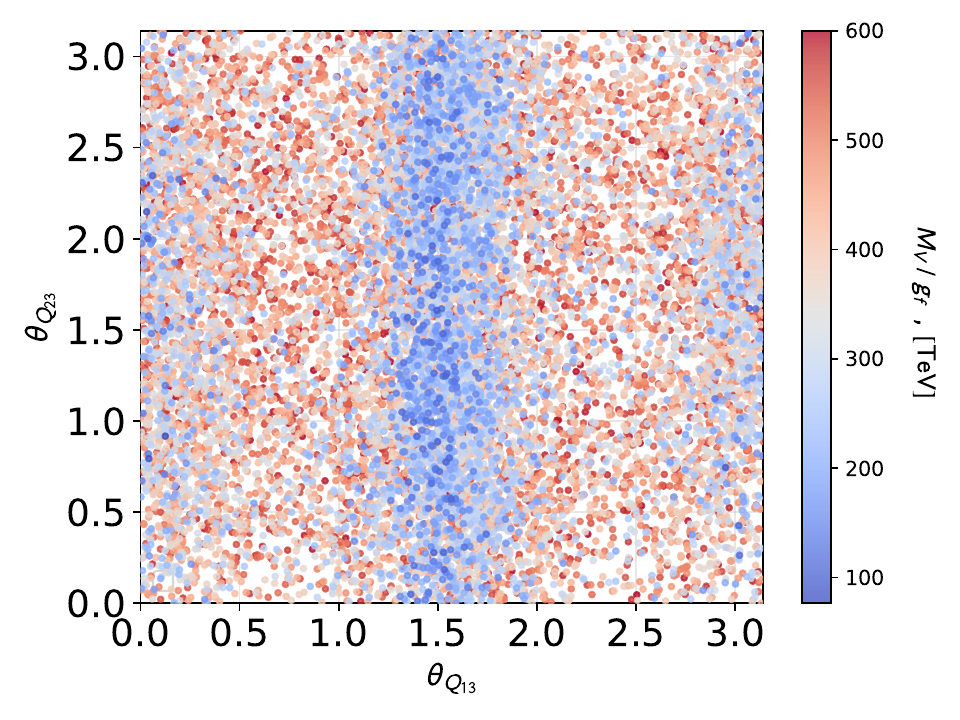}
}%
\caption{Overview of the distribution of the scan points passing the global fit within $1\sigma$ of the SM $\chi^2$ from \superiso~ as a function of the leptonic angles $\thLmutau$ and $\thLetau$ (a) and of the quark angles $\thQdb$ and $\thQsb$ (b). The colour coding refers to the ratio $M_V/g_f$ with blue points corresponding to lower NP scales. }
\label{fig:DistributionAngles}
\end{figure}

\vspace{0.2cm}
We illustrate the effect of the rotation angles on the constraints in Fig.~\ref{fig:DistributionAngles}, which shows all the points passing the combined flavour constraints at $2 \sigma$ in the global fit as a function of their flavour rotation angles in the quark ($\thQdb, \thQsb$) and lepton  ($\thLetau, \thLmutau$) sectors. The colour code indicates the ratio $M_V / g_f$ for each point with cooler colour indicating lower scales. As is quite clear from Fig.~\ref{fig:DistributionAngles} (a) and (b), points with precise flavour alignments (corresponding to vanishing spurions) are typically less constrained since they benefit from the global symmetry's protection discussed in Sec.~\ref{sec:SU2theo}. Additionally, smaller leptonic angles have a weaker correlation with the flavour constraints, reflecting the fact that, as shown in Table~\ref{tab:summaryintensity}, one of the dominant constraints arises from $K \to \pi \nu \nu$ processes. The residual preference for the $(13)_\ell$ or $(23)_\ell$ alignment (corresponding to $\thLmutau \sim \pi/2$ or $\thLetau \sim \pi/2$) reflects the constraints from the $K_L \to e \mu$ process. On the other hand, quark rotation angles show a clearer preference for the $(23)_Q$ and $(13)_Q$ alignments, which allow us to bypass completely the kaon-driven constraints and rely instead on weaker $B$-meson-driven limits. Note that the points' distribution is not symmetric in $\thQsb$ and $\thQdb$, reflecting the fact that the non-redundant parameterisation of Eq.~\eqref{eq:su2decomp} is asymmetric: first rotating $\thQdb$ by $\pi/2$ leads to the $(23)_Q$ flavour alignment, implying that the angle $\thQsb$ does not lead to a coupling with the first generation quark. On the other hand, this is not the case if we set $\thQsb \sim \pi/2$ since $V_x$ and $V_y$ do not commute in Eq.~\eqref{eq:su2decomp}.

\begin{figure}[t]
\centering
\subfloat[]{%
\includegraphics[width=0.47\textwidth]{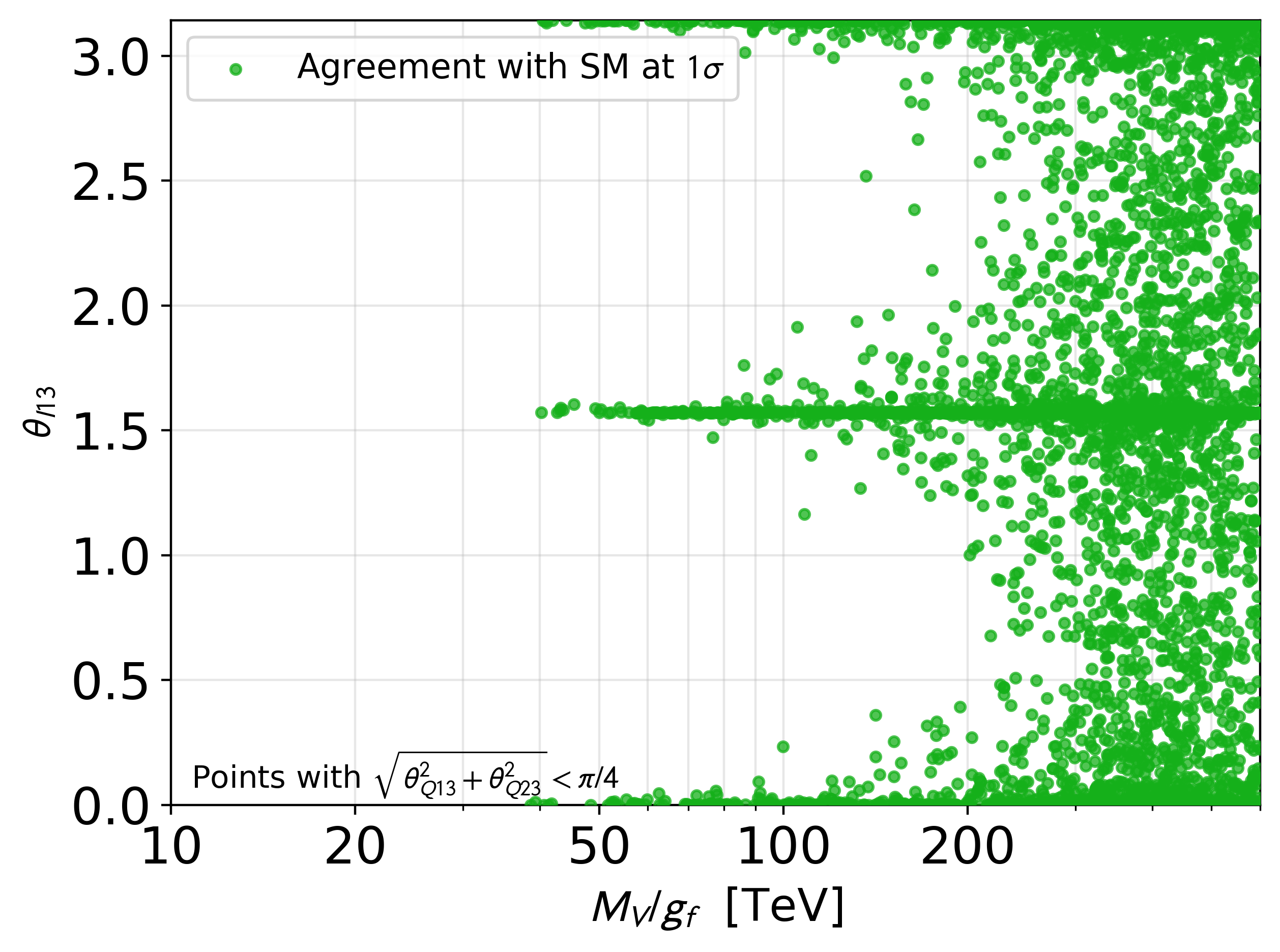}
}%
\subfloat[]{%
\includegraphics[width=0.47\textwidth]{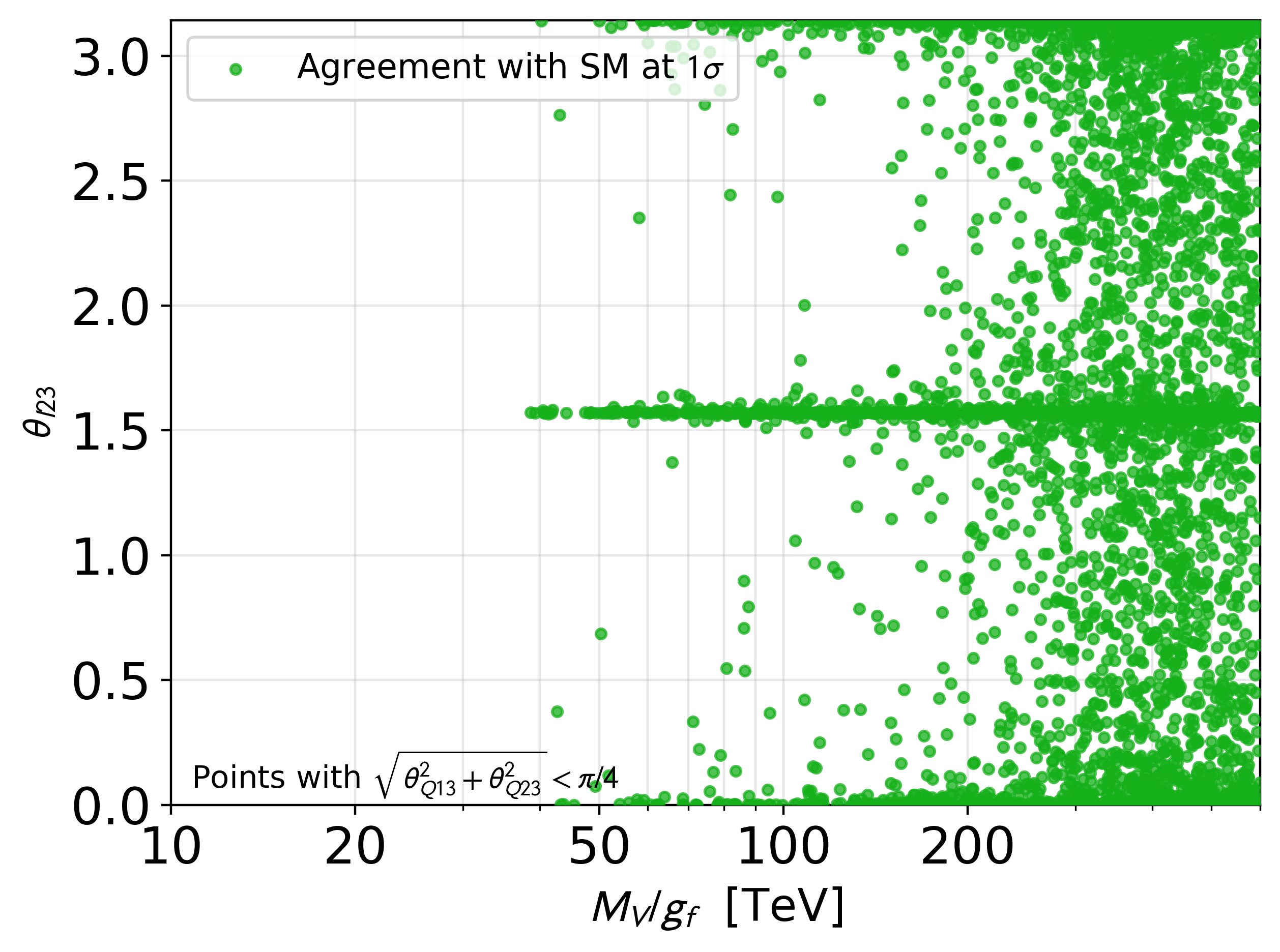}
}%
\\
\hspace{0.02\textwidth}
\subfloat[]{%
\includegraphics[width=0.47\textwidth]{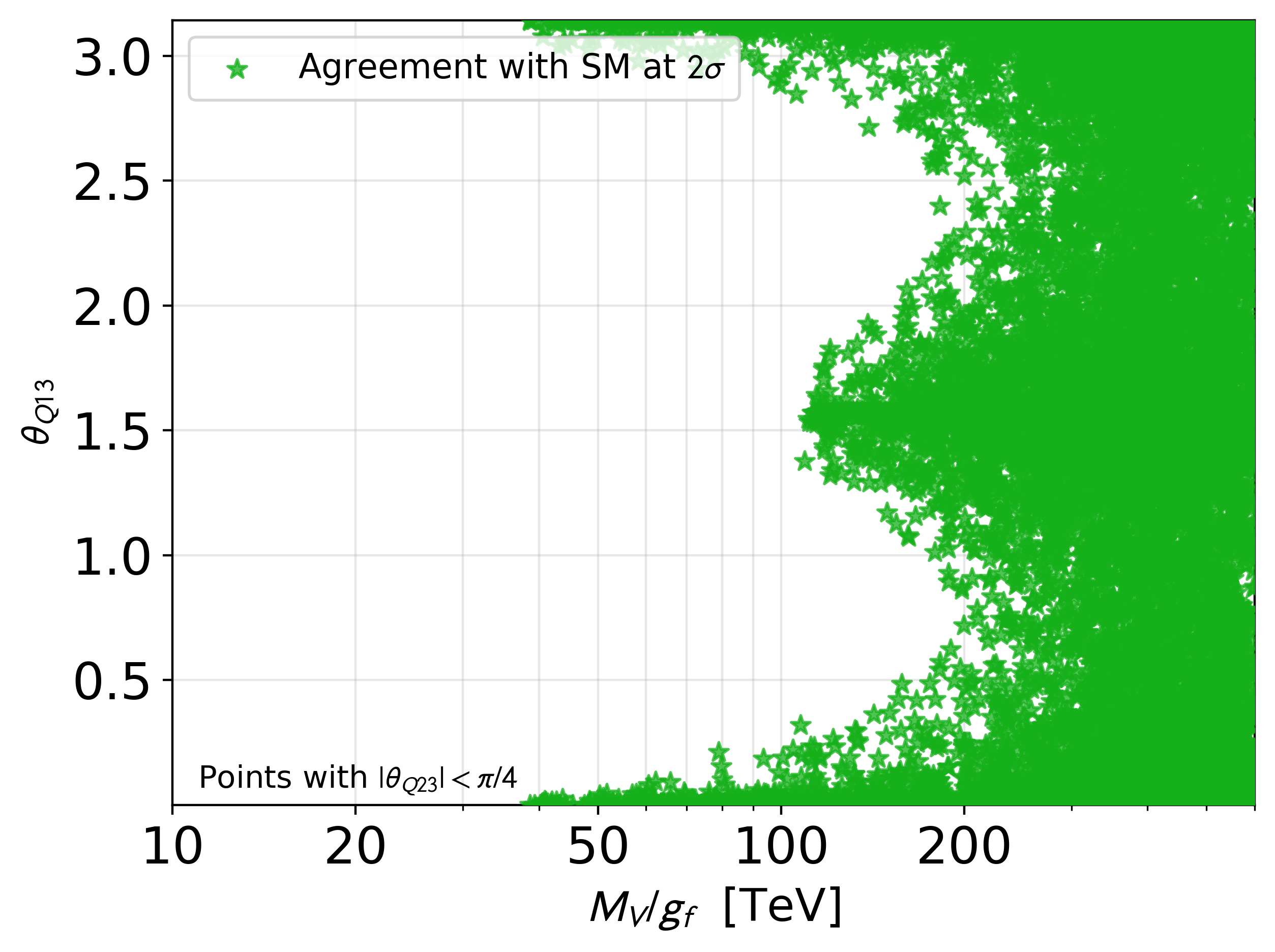}
}%
\subfloat[]{%
\includegraphics[width=0.47\textwidth]{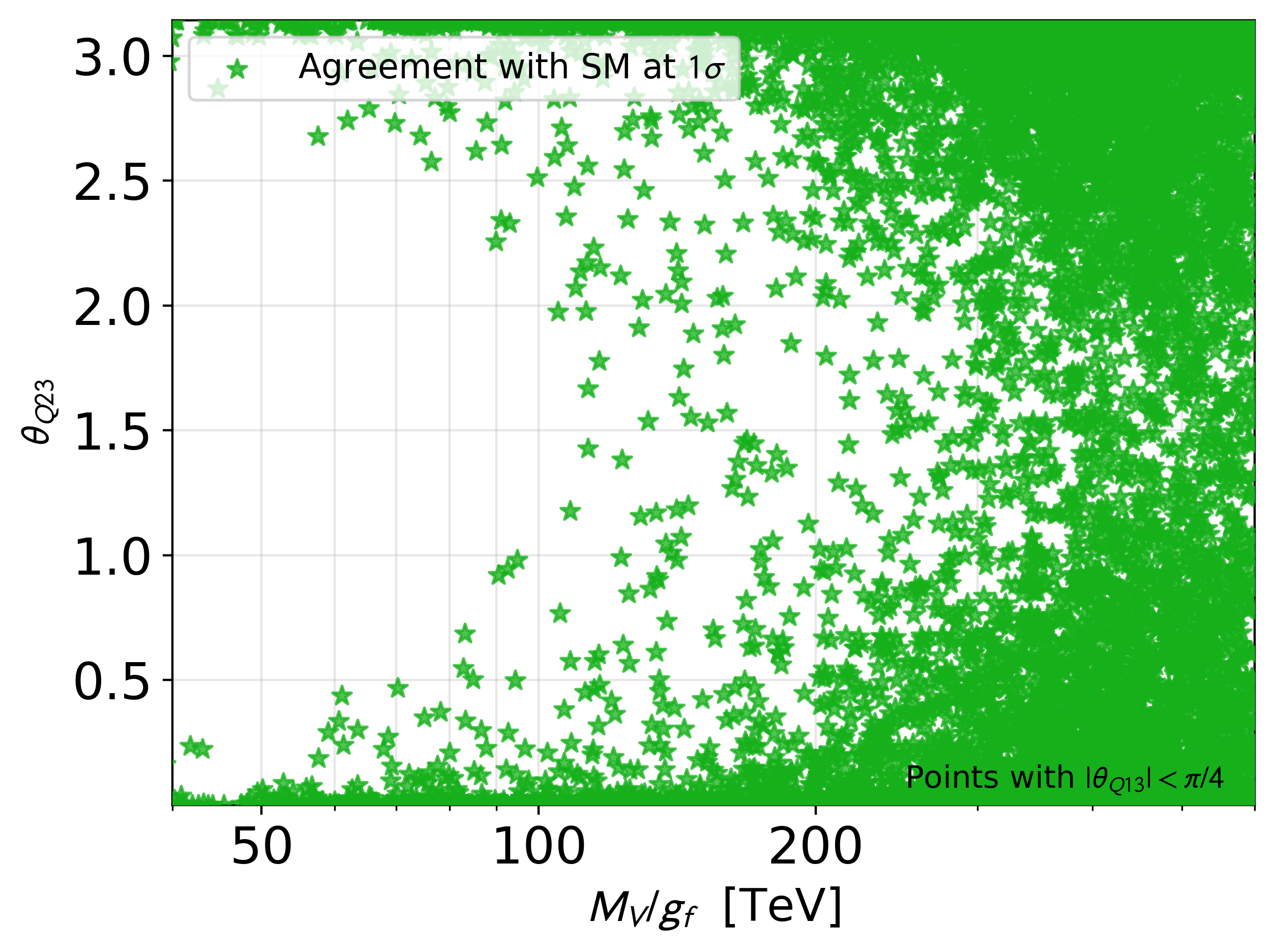}
}%
\caption{Parameter points passing all intensity frontier limits at $1\sigma$ in our global fit (green points) 
as a function of $M_V/ g_f$ on the x-axis and of the parameter: (a) $\theta_{\ell 13}$ , (b) $\theta_{\ell 23}$, (c) $\theta_{Q 13}$ , (d) $\theta_{Q 23}$ on the y-axis. We fix the quark rotation angles as indicated on the plots.}
\label{fig:ScanAnglesMass}
\end{figure}

We study further the parameters with the largest impact on the flavour constraints in Fig.~\ref{fig:ScanAnglesMass} by showing the distribution of points for each rotation angle as a function of the scale $M_V/ g_f$ (keeping only points where the other rotation angles are close to $0$). First of all, as can be seen in Figs.~\ref{fig:ScanAnglesMass} (a) and (b), the rotation angles $\thLetau$  (resp.  $\thLmutau$) between first (resp. second) and third generations leptons play an important role in the limits of the quark alignment $(12)_Q$. We trace this to the strong limits from  $\mu \to e$ transitions in Gold and $K_L \to e \mu$ decays. Note that the $\mu \to e$ has a residual dependence on the angle $\thLemu$, although since the transition rate is nuclei-dependent, measurements with different nuclei should be able to close this window~\cite{Davidson:2022nnl}.
For quark angles, Figs.~\ref{fig:ScanAnglesMass} (c) and (d) illustrate the importance of having small spurions (and thus definite flavour alignments as $(12)_Q$, $(13)_Q$ or $(23)_Q$ in order to reduce the constraints). \

\begin{figure}[t!]
\centering
        {\includegraphics[width=0.8\linewidth]{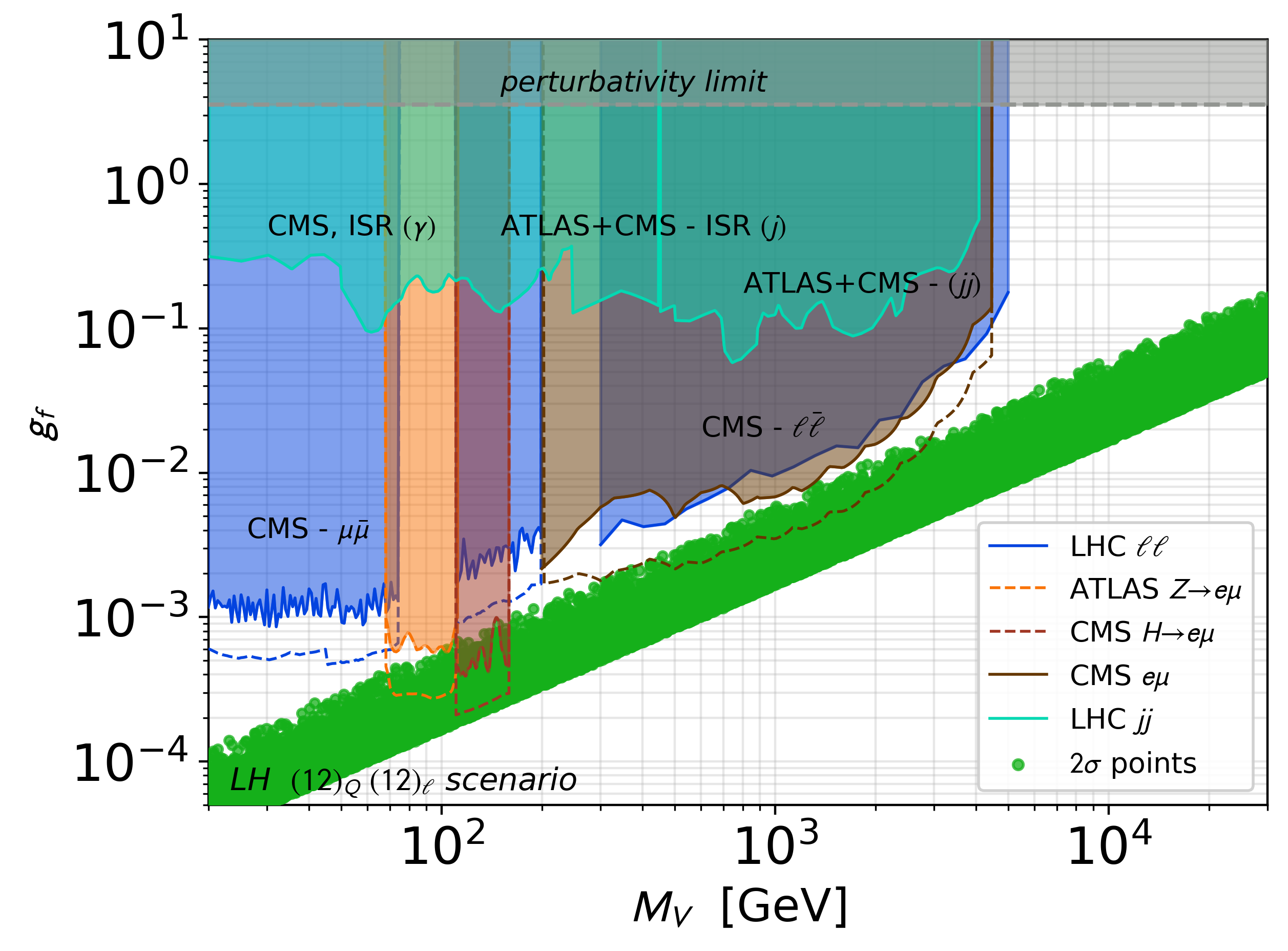}} \hspace{0.2cm}
        \caption{LHC and intensity frontier limits on our scenario (LH) with $12_Q$ and $12_\ell$ flavour alignments as a function of the scale $M_V / g_f$. Aquamarine regions indicate di-jets driven limits~\cite{ATLAS:2019fgd,CMS:2019gwf,ATLAS:2018qto,CMS:2016ltu,ATLAS:2019itm,CMS:2019emo,CMS:2019xai}, the blue regions are for di-muon limits~\cite{CMS:2019buh,CMS:2021ctt} and the brown exclusions from the LFV $e\mu$ search~\cite{CMS:2022fsw}. The orange and red exclusion regions at low mass are re-interpreted from LFV $H$ and $Z$ decay searches~\cite{ATLAS:2019old,ATLAS:2021bdj,CMS:2023pte}. Filled region indicates current constraints, while dashed lines are simple rescaling projections at HL-LHC full luminosity of 3~$\textrm{ab}^{-1}$.
        We overlaid the points passing all intensity frontier limits at $1\sigma$ in our global fit with the cut $M_V / g_f < 600 $ TeV.}
\label{fig:LHC_1212}
\end{figure}

\begin{figure}[t!]
\centering
\subfloat[]{%
\includegraphics[width=0.47\textwidth]{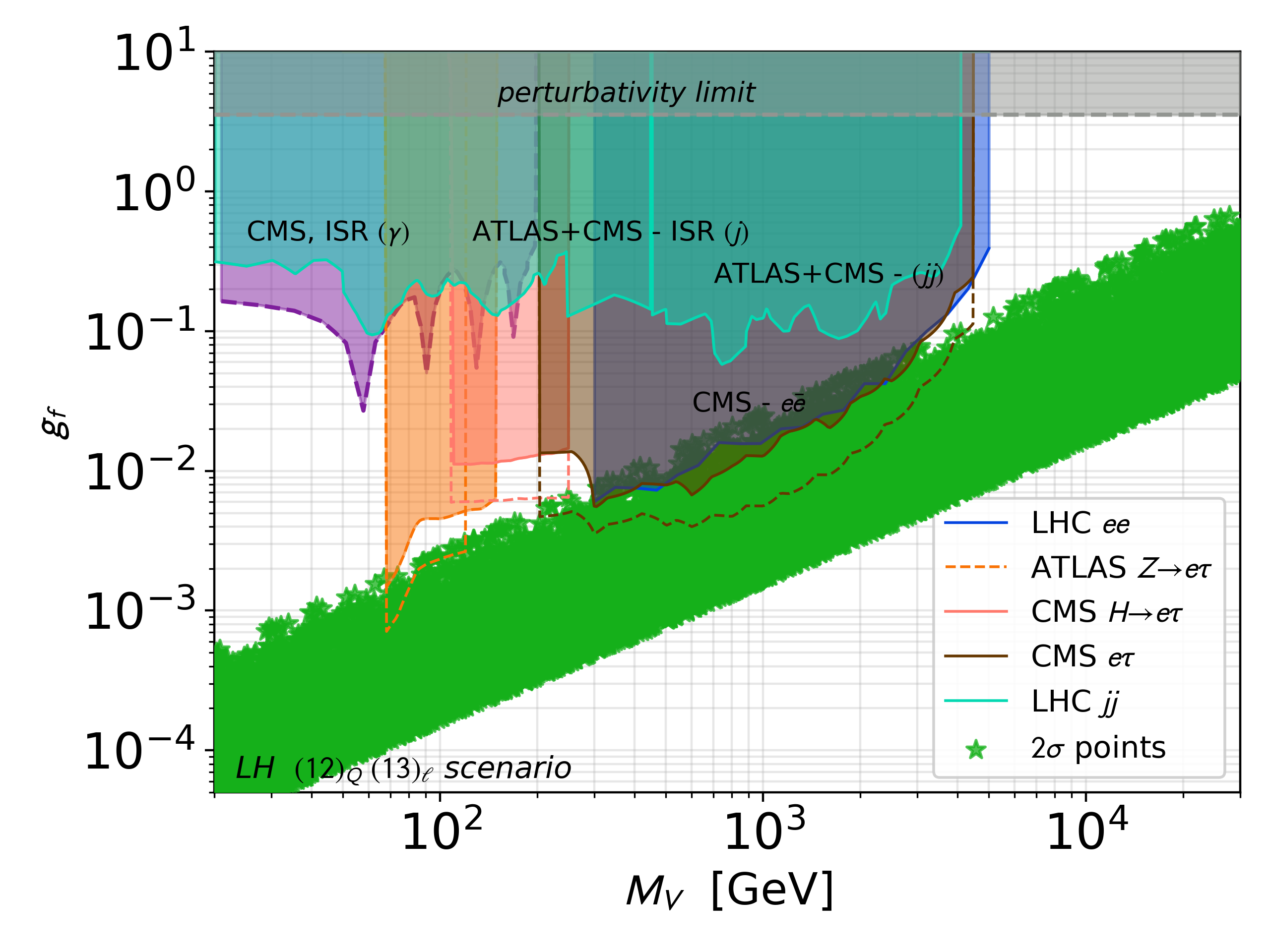}
}%
\hspace{0.02\textwidth}
\subfloat[]{%
\includegraphics[width=0.47\textwidth]{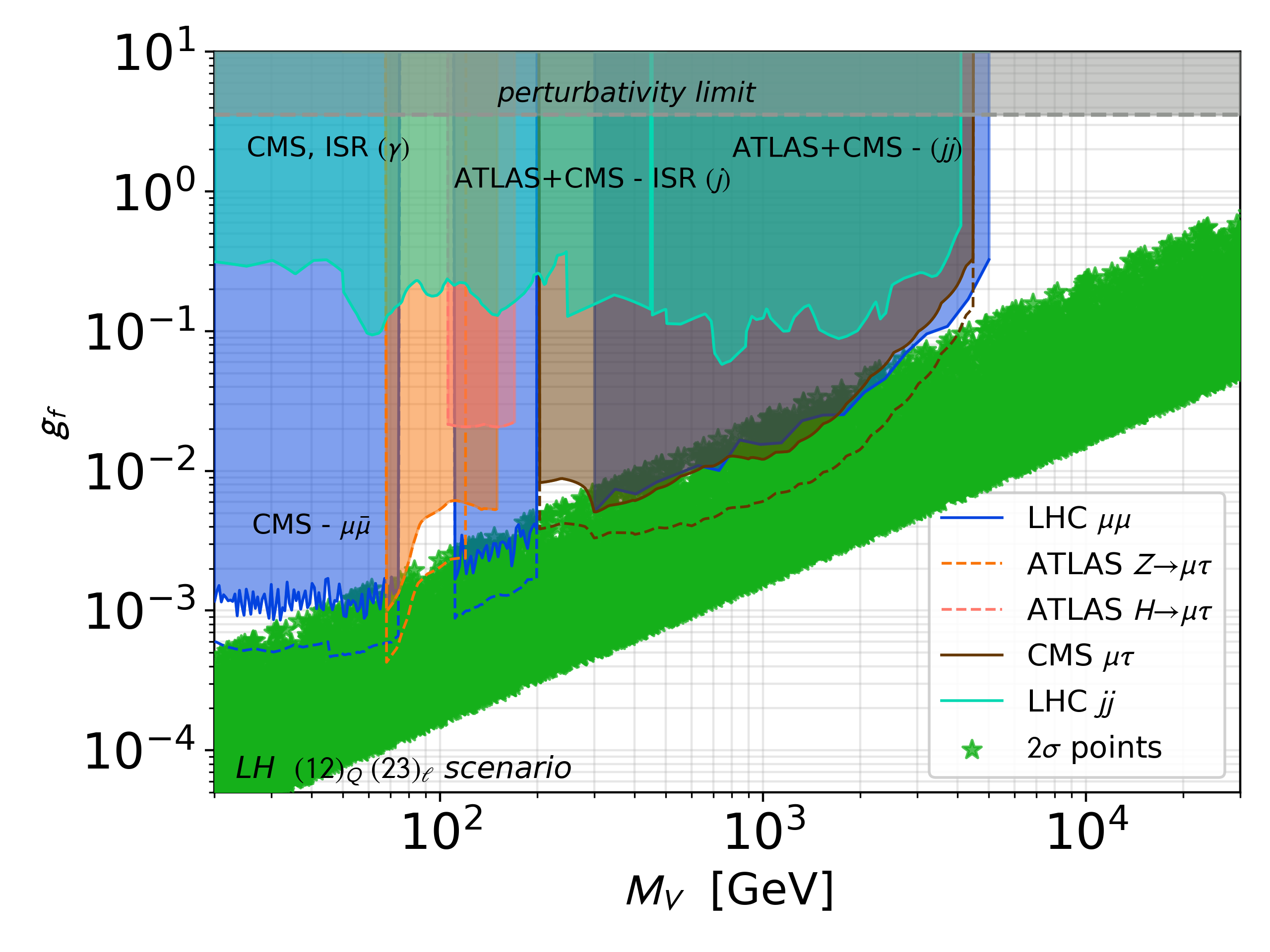}
}%
\subfloat[]{%
\includegraphics[width=0.47\textwidth]{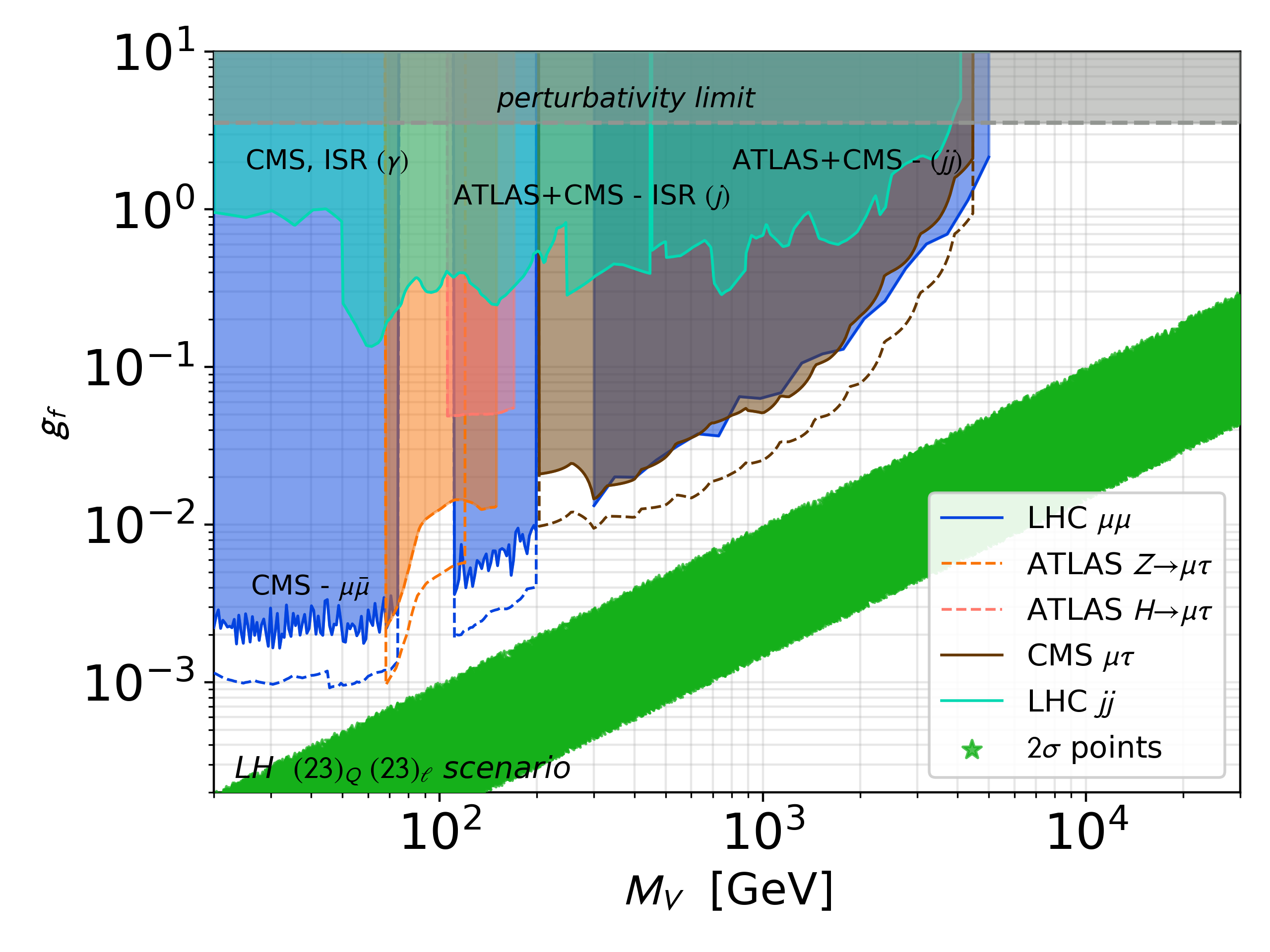}
}%
\caption{Same as in the previous plots, but with (a) $(12)_Q$ and $(13)_\ell$ flavour alignment  (b) $(23)_Q$ and $(23)_\ell$ flavour alignment.
        We overlaid the points passing all intensity frontier limits at $1\sigma$ in our global fit with the cut $M_V / g_f < 600 $ TeV. }
\label{fig:LHC_all}
\end{figure}

We finish this section by combining all the limits considered in this work, as shown in Fig.~\ref{fig:LHC_1212} and Fig.~\ref{fig:LHC_all}. The area below the green density of points which are within $1\sigma$ of the SM in our global fit is fully allowed and simply corresponds to the limit in our scan parameters of $M_V / g_f < 600$ TeV. We have restricted the rotation angles for each plot to retain the scan points close to a definite flavour alignment. The best prospects arise from the simplest flavour alignment $(12)_Q (12)_\ell$, for which the LHC searches are either with LFV final states (and in particular from $H$ and $Z$ LFV decay searches), or with di-leptons. 
For the case of the other flavour alignments, the relevance of LHC limits with respect to current flavour constraints depends mostly on the quark flavour alignment. In particular, for the $(23)_Q$ alignment, the production rates are severely limited at LHC which in turn implies that the corresponding constraints are typically weaker than the $B$-physics or $D^0$-oscillation ones.  

At low masses below the electroweak scale, the presence of LHC constraints depends mostly on the availability of di-muon final states, but can probe very small gauge couplings $g_f$ at the $10^{-3}$ level. Remarkably, we confirm the results from the previous sections that the LFV searches, including $e \tau$ final states can lead to limits on par or stronger than flavour physics constraints.
Finally, in order to provide an estimate of the potential of collider searches during the HL-LHC era, we give in Fig.~\ref{fig:LHC_1212} and Fig.~\ref{fig:LHC_all} simple projections for a 3~$\textrm{ab}^{-1}$ luminosity as dashed lines. These projections are based on the expected limits from the LHC searches described above re-scaled to 3~$\textrm{ab}^{-1}$, assuming that the dominant errors are statistical. We stress that dedicated searches, in particular at low masses for the $\tau$ LFV final states and at very low masses for the $e \mu$ LFV signatures, have the potential to lead to even stronger improvements. 

\section{Conclusion}
\label{sec:conclusion}

We have presented in this work a study of new  horizontal gauge groups $\suX$, acting on the SM fermions according to their generations. In the absence of any new additional fermions, there are only 6 sets of SM fermions that can lead to new anomaly-free gauge groups. Thus, depending on the choice of fermionic representations (doublets or triplets), there are only $20$ possibilities which we listed. We have then explored in detail the case of fermions in doublet representations of $\suX$.

This new gauge group will be broken by a scalar structure at scales around the $100$ TeV which must be described explicitly in full high energy theories. Since the new horizontal gauge coupling is a genuinely free parameter, fully independent from the SM gauge couplings even in the framework of a Grand Unified Theory, it can take a priori small values. Correspondingly, the new gauge bosons can obtain masses significantly smaller than the actual $\suX$ gauge breaking scale, and thus be within reach of a direct, on-shell, production at LHC. For the physics at LHC and in flavour experiments, the $\suX$ breaking sector can be solely parameterised in terms of spurions entering the mass matrices of the SM fermions. In this work, we thus explored their effects in terms of the mixing angles of the rotations to the fermion mass basis.
We have presented a complete study of the current LHC limits on the $6$ possible scenarios with SM fermions in doublets of $\suX$, emphasising the effect of the flavour alignments on the constraints based on the left-handed $\suX$ (LH) scenario.

We found that LHC has an excellent potential to test these scenarios, mainly due to the fact that the new gauge bosons interact in most scenarios with quarks (boosting production rates) and with leptons (allowing for efficient detection strategies at LHC). While di-lepton searches play an important role in constraining the low-mass region, lepton-flavour violating final states $e \mu$, $e \tau$ and $\mu \tau$ are shown to provide the dominant constraints in the rest of the parameter space. 
We identify in particular LFV signatures at low masses (around and below the Higgs boson and $Z$ boson masses) as being extremely sensitive to the new physics scenarios considered in this work. We have re-interpreted the latest ATLAS and CMS results with this final state since (with the exception of the latest CMS $e \mu$ search) the analysis had focused only on decays from the SM $Z$ and $H$ bosons. Future works in this direction, including in particular direct searches for new boson LFV decays beyond the $Z$ and the $H$ could lead to significant improvements in sensitivity for NP below the electroweak scale.

In a second time, we concentrated on flavour observables and intensity frontier experiments and the corresponding constraints on $\suX$ scenarios. Our main finding is that the $\suX$ symmetries provide a strong level of protection against several standard flavour-violating observables, including in particular meson oscillations and rare decays. The patterns of flavour violations predicted by the models are unique, in particular strongly distinct from Minimal Flavour Violation sequences of relevant flavour-violating observables. Indeed, we have shown that $\suX$ models do not predict flavour violation but rather flavour ``transfer'' between fermionic sectors, with processes such as $s \to d \mu \bar e$ being unsuppressed while others such as $\bar s d \to s \bar d$ being instead very strongly reduced. We thus have identified ``golden modes'' which are typically predicted  for the various models, and mostly rely on flavour transfer between the quark and the lepton sectors.

We finally studied in detail the left-handed scenario, corresponding to a $\suX$ gauge group acting on the SM $Q_L$ and $L$ doublets. In practice, we have included two generations of left-handed quarks and left-handed leptons in two $\suX$ doublets. We explored explicitly the relevant constraints, then studied them in two different ways. First, semi-analytically, using the expansion in terms of $\suX$ spurions for each observable, leading to the LHC constraints presented in Sec.~\ref{sec:LHC}. In a second time, we implemented all the relevant observables in the code \texttt{SuperIso} and numerically explored the accessible parameter space. In both cases, we confirm that the LHC is able to probe part of the parameter space of such flavourful NP models beyond current flavour constraints, particularly in the case of small $\suX$-breaking spurions. 

\acknowledgments
The authors thank Niels Fardeau for interesting discussions and insights on the leptonic effective operators, and Siavash Neshatpour for helping to implement the meson mixing observables.
L.D. is  supported by the European Union’s Horizon 2020 research and innovation programme under the Marie Skłodowska-Curie grant agreement No 101028626 from 01.09.2021. A.D. acknowledges partial support from the National Research Foundation in South Africa.

\begin{appendices}

\section{Scan procedure}
\label{sec:appScan}
The final results are obtained using the  code \superiso~to obtain a global $\chi^2$ for all intensity frontier observables. We have used $213$ experimental observables (using in particular the angular observables for $b \to s \ell \ell$ processes, including correlations as implemented in \superiso). More precisely, we have included:
\begin{itemize}
    \item All $b \to s \ell \ell$ observables as already implemented as a function of the WET coefficients in the public version of \superiso, including the new results from LHCb~\cite{LHCb:2022qnv,LHCb:2022zom}. 
    \item  Standard kaonic observables based on the $s d \ell \ell$ interaction as described in~\cite{DAmbrosio:2022kvb}.
    \item  $\Delta M_{B_s}, \Delta M_{B_d}$, $D^0$ (via $x_D$) and $K$ (via $\epsilon$) mixing observables.
    \item LFV semi-leptonic observables: $B \to K \nu \nu, B \to K \ell \ell^\prime$ and $K \to \pi \ell \ell^\prime$, together with $K_L \to e \mu$.
    \item Rare leptonic processes: The three-body final states: $\tau \to eee$, $\tau \to \mu \mu \mu$, $\tau \to \mu ee$, $\tau \to e \mu \mu $, $\mu \to eee$; and the radiative decays $\mu \to e \gamma, \tau \to e \gamma$ and $\tau \to \mu \gamma$.
    \item Muon to electron transition rate in gold $\textrm{CR} (\mu^- \, \textrm{Au} \to e^- \,\textrm{Au})$.
\end{itemize}
The expected inputs are the effective coefficients in the WET basis, therefore we have matched the $\suX$ model for the (LH) scenario with the effective theory at the electroweak scale (as the main goal of this scan is to constrain the parameter space with LHC searches) using the results from Sec.~\ref{sec:flavourobs}. We note that this is straightforward as the dominant constraints arise from four-fermion operators that are generated directly at tree level.
In order to provide a thorough scan of the relevant parameter space, we have interfaced our \superiso~distribution with \bsmart~\cite{Goodsell:2023iac} which directs the scan based on the \multinest~\cite{Feroz:2008xx} routines to converge on phenomenologically relevant regions as determined by the \superiso~$\chi^2$ output.
Our input parameters are based on the parameterisation described in Eqs.~\eqref{eq:gaugefermionMass},~\eqref{eq:su2decomp} and~\eqref{eq:effcharge}. We do not include the complex phases in this analysis and parameterised the angles based on the down-quark basis and $(12)_Q (12)_\ell$ flavour alignment. Each angle is varied between $[-\pi,\pi]$, the mass $M_V$ is scanned in the range $[10,600]$ TeV (the upper limit is chosen to escape all flavour constraints regardless of the rotation angles) and we fix $g_f=1$ in the scan since it is a fully redundant parameter for the intensity frontier limits. The final dataset of points presented in Figs.~\ref{fig:LHC_1212} and ~\ref{fig:LHC_all} is thus obtained by assigning randomly to  the points from the scan a value of $g_f$ and of $M_V$ while keeping the original ratio $M_V / g_f$ fixed.

\section{Details on some flavour-violating processes}
\label{sec:appFV}

We collect in this appendix additional details on flavour observables that we used in the full numerical approach.

\paragraph{Heavy flavour LFV decays}

We first complete in this paragraph the study of LFV decays of mesons by considering briefly the $B \to K$ and $B \to \pi$ processes. The experimental status corresponds to relatively weak constraints (at C.L 90\%): 
\begin{align}
    &\textrm{BR} (B^+ \to K^+  \mu^+ \tau^-) < 4.5 \cdot 10^{-5} &&(\textrm{BaBar}~\textrm{\cite{BaBar:2012azg}}) \nonumber \\
    &\textrm{BR} (B^+ \to K^+  \mu^- \tau^+) < 2.8 \cdot 10^{-5} &&(\textrm{BaBar}~\textrm{\cite{BaBar:2012azg}}, \textrm{see also}~\textrm{\cite{LHCb:2020khb}}) \nonumber \\
        &\textrm{BR} (B^+ \to K^+  e^+ \mu^-) < 7.0 \cdot 10^{-9} &&(\textrm{LHCb}~\textrm{\cite{LHCb:2019bix}})  \\
    &\textrm{BR} (B^+ \to K^+  e^- \mu^+) < 6.4 \cdot 10^{-9} &&(\textrm{LHCb}~\textrm{\cite{LHCb:2019bix}}) \nonumber \\
    &\textrm{BR} (B^0 \to  e^\pm \mu^\mp) < 1.4 \cdot 10^{-5} &&(\textrm{LHCb}~\textrm{\cite{LHCb:2019ujz}}) \nonumber \\
    &\textrm{BR} (B_s^0 \to  e^\pm \mu^\mp) < 4.2 \cdot 10^{-5} &&(\textrm{LHCb}~\textrm{\cite{LHCb:2019ujz}}) \ .\nonumber
\end{align}

These types of decay have received a good level of attention in the last decade (see e.g.~\cite{Glashow:2014iga,Calibbi:2015kma,Crivellin:2015era,Becirevic:2016oho,Becirevic:2016zri,Guadagnoli:2018ojc,Descotes-Genon:2020buf}). Quantitative estimates for the decay rates rely on the lattice estimates of the form factors. These limits are typically subdominant when relevant (for the cases of $(23)_Q$, we considered the semi-analytical results from~\cite{Crivellin:2015era,Becirevic:2016zri}). For instance, in the case of a $(23)_Q(23)_\ell$ flavour orientation for the (LH) scenario, we obtain
\begin{align}
    \textrm{BR}(B^+ \to K^+  \mu^- \tau^+ ) = 2.7 \cdot 10^{-7} \left( \frac{10 \, \textrm{TeV}}{M_V/g_f}\right)^4 \ ,
\end{align}
which leads to a significantly weaker limit than what could be obtained from $D^0$ oscillations.

\paragraph{$B_s$ and $B_d$ mixing}
In general, the relevant four-quark currents (following the conventions of Ref.~\cite{Altmannshofer:2007cs}), for the case of $b-s$ mixing are:
\begin{align}
    \mathcal{O}_1 & = (\bar{s} \gamma^\mu P_L b) \, (\bar{s} \gamma^\mu P_L b)\, , &&& 
    \widetilde{\mathcal{O}}_1 & = (\bar{s} \gamma^\mu P_R b) \, (\bar{s} \gamma^\mu P_R b)\, , \nonumber\\
    \mathcal{O}_2 & = (\bar{s} P_L b) \, (\bar{s} P_L b)\, , &&&
    \widetilde{\mathcal{O}}_2 & = (\bar{s} P_R b) \, (\bar{s} P_R b)\, , \nonumber \\
    \mathcal{O}_4 & = (\bar{s} P_L b) \, (\bar{s} P_R b) \, ,
\end{align}
with similar operator definitions for the $D_0-\bar{D}_0$ mixing, obtained by replacing $s \to u$ and $b \to c$.

The $B_s$-mixing matrix element can be computed as a function of the bag parameters $B^{(i)}_{B_s}$, relative to each operator $\mathcal{O}_i$~\cite{FermilabLattice:2016ipl}:
\begin{align}
    \langle  B_s |\mathcal{O}_1  |  \bar B_s \rangle &=   \langle  B_s |\widetilde{\mathcal{O}}_1  | \bar B_s \rangle = \frac{2}{3} f_{B_s}^2 M_{B_s}^2  B^{(1)}_{B_s}\ , \\ 
    \langle B_s |\mathcal{O}_2  | \bar B_s \rangle &=   \langle  B_s |\widetilde{\mathcal{O}}_2  |  \bar B_s \rangle = -\frac{5}{12} \left( \frac{M_{B_s}}{M_b+M_s}\right)^2 f_{B_s}^2 M_{B_s}^2  B^{(2)}_{B_s},\ , \\
      \langle B_s |\mathcal{O}_4  | \bar B_s \rangle &= \frac{1}{2} \left[\left( \frac{M_{B_s}}{M_b+M_s}\right)^2+\frac{1}{6} \right]f_{B_s}^2 M_{B_s}^2  B^{(4)}_{B_s} \, .
\end{align}
The bag parameters must be run from the scale of the flavour gauge bosons to the $B_s$ scale. We then constrain the ratio $R_{\Delta M_s}$ between the SM prediction and the NP contribution to the mass difference of the neutral $B_s$ mesons following~\cite{ParticleDataGroup:2020ssz}:
\begin{equation}
 \Delta M_{B_s} / \Delta M^{\rm SM}_{B_s} =  \left| 1+ \sum_{i=1}^2  R_i \frac{C_i+\widetilde{C}_i }{C_{1}^{\rm SM}(\mu_b) } + R_4 \frac{C_4}{C_{1}^{\rm SM}(\mu_b) }  \right|\ ,
\end{equation}
where we have defined the ratios of bag parameters at the $b$ scale by
\begin{align}
R_i \, \equiv \, \frac{\langle B_s| \mathcal{O}_i(\mu_b)| \bar{B}_s \rangle}{\langle B_s| \mathcal{O}_1(\mu_b)| \bar{B}_s \rangle }\ ,
\end{align}
and $R_i = (0.85,1.16,1.073,0.990)$ from~\cite{FermilabLattice:2016ipl}.

\paragraph{$D_0$ mixing}
The $D_0$- mixing matrix can be obtained in a similar way, with the parameter of interest being $x_D$ defined as:
\begin{align}
    x_D ~\equiv~ \frac{\Delta M_{D_0}}{\Gamma_{D_0}} \ .
\end{align}
The effective operators in the LL, LR and RR basis are defined as (note that for the full \superiso~implementation we rely on the so-called SUSY basis, see e.g.~\cite{Aebischer:2020dsw}):
\begin{align}
    \mathcal{O}_1 & = (\bar{u} \gamma^\mu P_L c) \, (\bar{u} \gamma_\mu P_L c)\, ,  \nonumber\\
    \mathcal{O}_2 & = (\bar{u} \gamma^\mu P_R c) \, (\bar{u} \gamma_\mu P_L c)\, , \nonumber\\
    \mathcal{O}_6 & = (\bar{u}  \gamma^\mu P_R c) \, (\bar{u}  \gamma_\mu P_R c) \ ,
\end{align}
with $x_D$ then written as~\cite{Golowich:2007ka,Golowich:2009ii,Bause:2019vpr}:
\begin{align}
   & x_D ~\equiv~ -\frac{\langle D_0| \mathcal{O}_1(\mu_b)| \bar{D}_0 \rangle}{\Gamma_{D_0} M_{D_0}} \left[ r \, \mathcal{C}_1  + \sqrt{r} \mathcal{C}_3  R_2  + \frac{2}{3} (\sqrt{r}-\frac{1}{r^4}) \mathcal{C}_2  R_3   + r \, \mathcal{C}_6  \right] \ .
\end{align}
The matrix elements are given by:
\begin{align}
    \langle D_0| \mathcal{O}_1(\mu_b)| \bar{D}_0 \rangle = 0.0805 \ ,\\
    R_2 \equiv \frac{\langle D_0| \mathcal{O}_2(\mu_b)| \bar{D}_0 \rangle }{\langle D_0| \mathcal{O}_1(\mu_b)| \bar{D}_0 \rangle } = -2.57 \ ,\\
    R_3 \equiv \frac{\langle D_0| \mathcal{O}_3(\mu_b)| \bar{D}_0 \rangle }{\langle D_0| \mathcal{O}_1(\mu_b)| \bar{D}_0 \rangle } = 3.41 \ ,
\end{align}
and the strong coupling renormalisation factor $r$ with $\mu = 3$ GeV:
\begin{align}
    r  = \left( \frac{\alpha_s (M_V)}{\alpha_s (m_t)} \right)^{2/7}  \left( \frac{\alpha_s (m_t)}{\alpha_s (m_b)} \right)^{6/23} \left( \frac{\alpha_s (m_b)}{\alpha_s (\mu)} \right)^{6/25} \ .
\end{align}

We can put limits directly from the experimental average~\cite{ParticleDataGroup:2020ssz},
\begin{align}
    x_D^{\rm exp} = (4.09 \pm 0.48) \times 10^{-3} \ ,
\end{align}
 noting that there is no definite SM prediction yet.

For our illustrative (LH) scenario with $12_Q$, we obtain at leading spurion order:
 \begin{align}
     x_D = 3.7 \cdot 10^{-4} \ \thQdb^2 \times \left( \frac{100 \, \textrm{TeV} }{M_V/g_f} \right)^4 \ .
 \end{align}

\end{appendices}

\bibliographystyle{JHEP2}
\bibliography{bibliography}

\end{document}